\documentclass{article}
 \usepackage{epsfig}
 \usepackage{amsmath}
 \usepackage{graphics}
 \usepackage{latexsym}
 \usepackage{amssymb}

\def\vb{\mathbf{v}}
\def\p{\partial}
\def\ubar{\bar{u}}
\def\abar{\bar{a}}
\def\rhobar{\bar{\rho}}
\def\mbar{\overline{\rho u}}
\def\ub{\mathbf{u}}

\def\xb{\mathbf{x}}

\newcommand{\eref}[1]{(\ref{#1})}

\def\spacce#1{\hskip #1pt}
\def\drawline#1#2{\raise 2.5pt\vbox{\hrule width #1pt height #2pt}}
\def\solid{\drawline{24}{.5}\nobreak}
\def\bdash{\hbox{\drawline{4}{.5}\spacce{2}}}
\def\dashed{\bdash\bdash\bdash\bdash\hskip-2pt\nobreak}
\def\bdot{\hbox{\drawline{1}{.5}\spacce{2}}}

\def\dashdot{\bdash\bdot\bdash\bdot\bdash\bdot}

 \newtheorem{theorem}{Theorem}[section]
 \newtheorem{lemma}[theorem]{Lemma}

 \author{Greg Norgard   \thanks{Graduate Student,Department of
         Applied Mathematics, University of Colorado,
         Boulder, Colorado, 80309, US.}
        \and Kamran Mohseni \thanks{Associate Professor of
        Aerospace Engineering Sciences; Affiliated faculty
        in the Applied Mathematics Department, University of
        Colorado, Boulder, Colorado, 80309, US.}
        }

\title{A New Regularization of the One-Dimensional Euler and Homentropic Euler Equations}

\begin{document}
\maketitle

\begin{abstract}
This paper examines an averaging technique in which the nonlinear
flux term is expanded and the convective velocities are passed
through a low-pass filter. It is the intent that this modification
to the nonlinear flux terms will result in an inviscid
regularization of the homentropic Euler equations and Euler
equations.  The physical motivation for this technique is presented
and a general method is derived, which is then applied to the
homentropic Euler equations and Euler equations.  These modified
equations are then examined, discovering that they share the
conservative properties and traveling wave solutions with the
original equations.  As the averaging is diminished it is proven
that the solutions converge to weak solutions of the original
equations. Finally, numerical simulations are conducted finding that
the regularized equations appear smooth and capture the relevant
behavior in shock tube simulations.
\end{abstract}

\section{Introduction}
This paper examines a modification of the homentropic Euler and
Euler equations where flux term is expanded and the convective
velocities are passed through a low-pass filter.  It is the intent
that this modification to the nonlinear flux terms will result in
an inviscid regularization of the homentropic Euler equations and
Euler equations.  The ultimate goal with this regularization is to
develop a proper modeling of small scale behavior so that both
shocks and turbulence can be captured in one comprehensive
technique.  This paper primarily addresses the shock capturing
capabilities of the developed technique.

The inspiration for this new technique comes from recent work done
on a similarly modified Burgers equation.  The Burgers equation,
$u_t+uu_x=0$, is considered a simplistic model of compressible flow
which forms shocks readily.  Classically, this equation is
regularized with a dissipative term, such as viscosity or
hyper-viscosity.  Additionally Burgers equations can be regularized
with linear dispersion resulting in the KdV equations. Recently, work has
been done by our group and others, investigating a regularization of
the Burgers equations where the nonlinear term is manipulated by
averaging the convective velocity
\cite{Mohseni:06l,Norgard:08b,BhatHS:06a,BhatHS:08a,NorgardG:08a,Holm:03a},
\begin{subequations}\label{CFB}
\begin{eqnarray}
u_t +\ubar u_x=0\\
\ubar=g^\alpha \ast u\\
g^\alpha=\frac{1}{\alpha} g(\frac{x}{\alpha}),
\end{eqnarray}
\end{subequations}
where $g$ is an averaging kernel with emphasis on the Helmholtz
filter.  In this paper we extend this technique to the homentropic
Euler and Euler equations.

The technique used on the Burgers equations has been thoroughly
investigated and been established as a valid shock regularization
technique.  It has been shown that solutions to Equations
(\ref{CFB}) exist and are unique \cite{Norgard:08b,BhatHS:06a} for a
wide variety of filters.  This result has also been extended for a
multiple dimensional version of the equations as well
\cite{Norgard:08b}.  It has been shown that when the initial
conditions are $C^1$, the solution  remains $C^1$ for all time
\cite{Norgard:08b}.  Furthermore, for bell shaped initial conditions
it is proven that as the averaging approaches zero ($\alpha \to 0$),
the solutions to Equation (\ref{CFB}) converge to the entropy
solution of inviscid Burgers equation \cite{NorgardG:08a}. There is
also convincing evidence that this holds true for all continuous
initial conditions \cite{NorgardG:08a}. These analytical results
along with multiple numerical results regarding shock thickness and
energy decay results show that this is a valid shock regularization
technique.

The work on the regularization of the Burgers equation is inspired
by and related to work done on the Lagrangian Averaged Navier-Stokes (LANS-$\alpha$) equations
\cite{Holm:04a,Foias:01a,Marsden:01h,Mohseni:03a,Mohseni:05e,Marsden:98b,Chen:98a,IlyinAA:06a}.
The LANS-$\alpha$ equations also employ an averaged velocity in the
nonlinear term and have been successful in modeling some turbulent
incompressible flows.

It is thought that a regularization could be accomplished
for the equations that describe compressible flow. Encouraged by the
results for Burgers equation, the next step is to attempt to
introduce averaging into the one-dimensional homentropic Euler
equations, a simplified version of the full Euler equations, where
pressure is purely a function of density. There have been several
attempts at such a regularization.

Inspired by the existence uniqueness proofs from the averaged
Burgers equations \cite{Norgard:08b,BhatHS:06a}, Norgard and Mohseni
\cite{NorgardGJ:09b}  averaged the characteristics of the
homentropic Euler equations to derive the equations
\begin{subequations}
\label{CAHE}
\begin{eqnarray}
\label{CAHEa}
\rho_t+\ubar \rho_x +\rho \frac{\abar}{a} u_x=0\\
\label{CAHEb}
u_t+ \ubar u_x+\frac{a\abar}{\rho}\rho_x=0\\
\label{CAHEc}
\ubar=g \ast u\\
\label{CAHEd} \abar=g \ast a
\end{eqnarray}
\end{subequations}
with $a^2=\gamma \rho^{\gamma-1}.$ While these equations were proven
to have a convenient existence and uniqueness proof, the equations
were ultimately found to have significant departures in behavior
from the homentropic Euler equations, specifically for the Riemann
problem.  What was discovered is that when the characteristics are averaged, there will be no creation or destruction of characteristics.  A shock in homentropic Euler equations can produce new characteristics in one of the Riemann invariants.  The new equations does not capture this behavior and results in departures from the desired behavior.

Using a Lagrangian averaging technique Bhat and Fetecau
\cite{BhatHS:06b} derived the following equations
\begin{eqnarray}
\rho_t+(\rho u)_x=0\\
w_t+(uw)_x-\frac{1}{2}( u^2 + \alpha^2 u_x^2)_x=-\frac{p_x}{\rho}\\
\rho w=\rho v-\alpha^2 \rho_x u_x\\
v=u-\alpha^2 u_{xx}.
\end{eqnarray}
While the solutions to the system remained smooth and contained much
structure it was found that the equations were ``not well-suited for
the approximation of shock solutions of the compressible Euler
equations.''

Another attempt by Bhat, Fetecau, and Goodman used a Leray-type
averaging  \cite{BhatHS:07a} leading to the equations
\begin{eqnarray}\label{LerayAveragedHomentropicEuler}
\rho_t+\ubar \rho_x +\rho u_x=0\\
u_t + \ubar u_x + \frac{p_x}{\rho}=0\\
u=\ubar-\alpha^2 \ubar_{xx}.
\end{eqnarray}
with $p=\kappa \rho ^\gamma$.  They then showed that weakly
nonlinear geometrical optics (WNGO) asymptotic theory predicts the
equations will have  global smooth solutions for $\gamma =1$ and
form shocks in finite time for any other value of  $\gamma$; namely $\gamma \neq 1$.

Additionally in 2005, H. S. Bhat et. al. \cite{Mohseni:05d} applied
the Lagrangian averaging approach to the full compressible Euler
equations.  Their approach successfully derived a set of
Lagrangian Averaged Euler (LAE-$\alpha$) equations.
However, these equations were quite long and complicated, and it seemed that numerical simulations involving these equations would be impractical for real world applications.

This paper examines another attempt at developing a regularization
of the homentropic Euler and Euler equations with positive results
thus far.  Section \ref{Motivationsection} details the motivation
behind the technique with section \ref{generalmethodsection}
detailing the general method.  Section \ref{filtersection} specifies
the averaging kernels that we consider in the regularizations.
 Section \ref{NMHEsection} examines the modified homentropic Euler
equations looking at conservation of mass and momentum, traveling
wave solutions, eigenvalues, convergence to weak solutions of the
original homentropic Euler equations, and finally examining some
numerical results.  Section \ref{NMEsection} examines the same
properties, but with the modified Euler equations. All is then
followed by concluding remarks.

\section{Motivation}\label{Motivationsection}
\begin{figure}[!ht]
\begin{center}
\begin{minipage}{0.48\linewidth} \begin{center}
  \includegraphics[width=.9\linewidth]{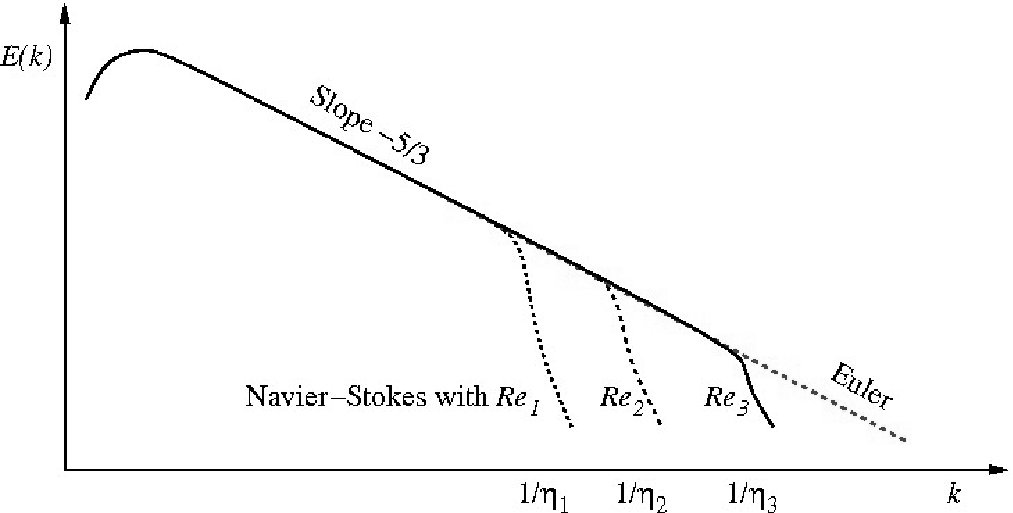}
\end{center} \end{minipage}
\begin{minipage}{0.48\linewidth} \begin{center}
  \includegraphics[width=.9\linewidth]{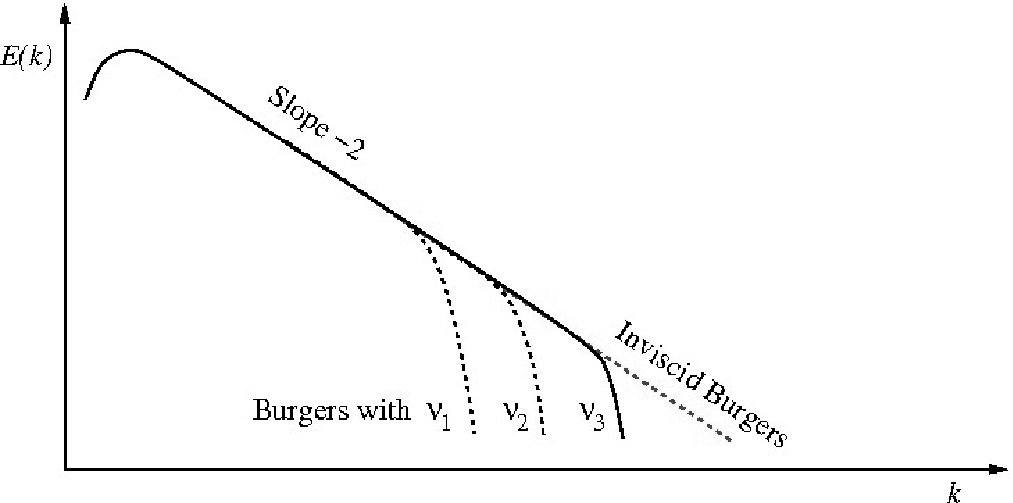}
\end{center} \end{minipage}\\
\begin{minipage}{0.48\linewidth}\begin{center} (a) \end{center} \end{minipage}
\begin{minipage}{0.48\linewidth}\begin{center} (b) \end{center}
\end{minipage}
\caption{Schematics of energy cascade for the Navier-Stokes/Euler
equations and the viscous/inviscid Burgers equation are shown. In an
inviscid flow, both turbulence and shocks  show continuous
generation of higher wave modes indefinitely. (a) Energy cascades
from high wavelengths to lower wavelengths at a predicted rate of
-$\frac{5}{3}$ for the Navier-Stokes/Euler equations.  For the
Navier-Stokes equations, the kinetic energy drops drastically upon
reaching a certain wavelength; the Kolmogorov scale, $\eta$.  For
the Euler equations the cascade continues on indefinitely.  Here
$Re_1 < Re_2 <Re_3$. (b) For the Burgers equation, shocks can form
from continuous initial conditions.  The energy cascade has a slope
of -$2$ when shocks form, until viscosity begins to exert its
influence and balance the steepening effect of the nonlinear term.
Here $\nu_3<\nu_2<\nu_1$.} \label{kscale}
\end{center}
\end{figure}

The motivation for our technique stems from a simple concept:  that
nonlinear terms generate high wave modes as continuously as time
progresses. Consider the mechanics behind shock formation and
turbulence. The nonlinear convective term $\ub\cdot\triangledown
\ub$ generates high wave modes, by transferring energy into smaller
scales as time progresses.  This nonlinear term is found in the
Burgers equations where it causes nonlinear steepening resulting in
shocks.  It is also found in the Euler and Navier-stokes equations
where is generates high wave modes by tilting and stretching
vortices \cite{BradshawP:76a}.  Thus this nonlinear term is
cascading energy down into smaller and smaller scales. In the 3D
Navier-Stokes and 3D Euler equations, the energy cascade has a slope
of -$\frac{5}{3}$ until the Kolmogorov scale, illustrated in figure
\ref{kscale}a.  The Burgers equation has an energy cascade slope of
-$2$ until viscosity begins to dominate; seen in figure
\ref{kscale}b \cite{Kraichnan:68a, GurbatovSN:97a}.  It is by
reducing this cascade of energy after some scale that we intend to
regularize the Euler equations. In the Burgers equation the
nonlinear term $u u_x$ was replaced with $\ubar u_x$.  A low pass
filtered velocity will have less energy in its high wave modes after
the scale $\frac{1}{\alpha}$. Thus when inserted into the nonlinear
term, the energy cascade will be lessened.  This modification to the
Burgers equations was found to result in a regularization of the
equations. It is our hypothesis that a similar modification to the
Euler equations will have similar results.

\section{Averaging kernels}\label{filtersection}
In our previous work, \cite{Norgard:08b,NorgardG:08a} the low pass
filters were assumed to have certain properties.  Specifically the
convolution kernels $g$ were assumed to be normalized, nonnegative,
decreasing, and even.  This is summarized in table
\ref{FilterProperties}.  The physical rational behind these
properties can be found in our previous work \cite{Norgard:08b}. For
this paper similar properties are assumed for the filters.  However,
only sections \ref{generalmethodsection} and
\ref{generalconservationsection} deal with a general filter. All
other sections deal exclusively with the Helmholtz filter which
possesses all these characteristics.  The Helmholtz filter is
defined as
\begin{equation}\label{helmholtzfilter}
u=\ubar-\alpha ^2 \ubar_{xx},
\end{equation}
and thus has an averaging kernel of
\begin{equation}\label{helmholtzkernel}
g^\alpha(x)=\frac{1}{2\alpha}\exp(-\frac{|x|}{\alpha}).
\end{equation}

\begin{table}
\caption{\label{FilterProperties}This table succinctly lists the
requirements of the low pass filters employed.}
\begin{center}
\begin{tabular}{|l|c|}
\hline
Properties & Mathematical Expression\\
\hline
Normalized & $\int g$=1 \\
Nonnegative & $g(\xb) >0,\, \forall \xb $   \\
Decreasing & $|\xb_1|\ge |\xb_2| \Rightarrow g(\xb_1) \le g(\xb_2)$  \\
Symmetric  & $|\xb_1|= |\xb_2| \Rightarrow g(\xb_1) = g(\xb_2)$    \\
\hline
\end{tabular}
\end{center}
\end{table}

\section{A general method} \label{generalmethodsection}
The successful regularization of the Burgers equations can be extended in several ways.  As mentioned in the introduction, we had previously interpreted the CFB equations as an averaging of the characteristics \cite{NorgardGJ:09b}.  Bhat, Fetecau, and Goodman  interpreted the regularization of the Burgers equation as an averaging of the convective velocity,  a Leray-type
averaging  \cite{BhatHS:07a}.  Both of these interpretations, when extended to the homentropic Euler equations, have not led to the desired regularization of homentropic Euler equations.  Here we examine yet another interpretation  of the CFB equations, based on a conservation law perspective. This approach addresses the cascade of energy
generated by the nonlinear terms. We then discuss how the technique used in regularizing the Burgers equation can be extend to a general technique to be used on
conservation laws.  Begin by looking at the inviscid
Burgers equation
\begin{equation}\label{differentlyscaledBurgers}
u_t+(uu)_x=0.
\end{equation}
The flux term represents that the quantity $u$ is flowing into a
control volume with velocity $u$. When the product rule is applied
to the flux term the result is
\begin{equation}
u_t+u u_x + u u_x=0.
\end{equation}
Each nonlinear term in the above equation results in steepening
waves which cascades energy into higher wave modes. It is this
energy cascade that produces shocks. In section
\ref{Motivationsection}, it was discussed how this cascade of energy
can be reduced by filtering the convective velocity. Thus the
non-differentiated term is passed through a low pass filter. The
resulting equation is
\begin{equation}
u_t+\ubar u_x + \ubar u_x=0,
\end{equation}
which has been referred to as the convectively filtered Burgers
(CFB) equations.  The portrayal of the CFB equations here differs
from Equation \eref{CFB} and in previous papers
\cite{Mohseni:06l,BhatHS:06a,BhatHS:08a,Norgard:08b,NorgardG:08a,Holm:03a}
by a factor of 2, but is identical under rescaling.

Now we extend this technique to a more general method.  Suppose
there is a single or multiple conservation laws of the form
\begin{equation}
Q_t+(Qu)_x=0.
\end{equation}
The proposed method is to apply the product rule to the nonlinear
term and then apply a filter to the non-differentiated quantities.
This results in
\begin{equation}
Q_t+\bar{Q}u_x+\ubar Q_x=0.
\end{equation}
We believe that this filtering of the non-differentiated quantities
will reduce the energy cascade and provide a regularization of the
conservation law.  It is this general method that we apply to  the
homentropic Euler and Euler equations.

\section{Conservation of conserved quantities} \label{generalconservationsection}
Before we begin the examination of the modified homentropic Euler
and Euler equations we will first examine how the modification of
the nonlinear term still preserves the original conserved
quantities. Again consider the more general conservation law
\begin{equation}
Q_t+(Qu)_x=0
\end{equation}
and its regularized version
\begin{subequations}
\begin{align}\label{generalmodifieda}
Q_t+\bar{Q}u_x+\ubar Q_x=0\\
\label{generalmodifiedb}
\ubar=g*u\\
\label{generalmodifiedc}
 \bar{Q}=g*Q.
\end{align}
\end{subequations}
 Integrate
$Q_t$ over the spatial domain and substitute Equations
\eref{generalmodifiedb} and \eref{generalmodifiedc} and the full
definition of convolution to obtain
\begin{equation} \frac{\p}{\p
t}\int Q(x) dx=-\int\int g(x-y) Q(y) u'(x) \,dy \,dx - \int\int
g(x-y) u(y) Q'(x) \,dy \,dx.
\end{equation}
Integrate by parts to obtain \begin{equation} \frac{\p}{\p t}\int
Q(x) dx=\int\int g'(x-y) \left(Q(y) u(x) + u(y) Q(x)\right) \,dy
\,dx.
\end{equation}
If $g$ is even, then $g'$ is odd and $g'(x-y)$ is anti-symmetric
over $y=x$.  Clearly $\left(Q(y) u(x) + u(y) Q(x)\right)$ is
symmetric over $y=x$ and thus by symmetry
\begin{equation}
\frac{\p}{\p t}\int Q(x) dx=0.
\end{equation}
From this it is determined the modified conservation law still
conserves the original conserved quantity.  Thus we see that the
proposed method conserves the quantities that the original
conservation laws were designed to preserve.  This property is
independent of the filter so long that it is even, which was a
requirement of section \ref{filtersection}.

\section{The one-dimensional homentropic Euler
equations}\label{NMHEsection}
Examine the homentropic Euler equations
\begin{subequations}
\label{homentropicEulerEquations}
\begin{align}
\label{homentropicEulerEquationsa}
\rho_t+(\rho u)_x=0\\
\label{homentropicEulerEquationsb} (\rho u)_t+ (\rho u u+P)_x=0,
\end{align}
\end{subequations}
with  pressure defined as $P=\rho^\gamma$.  Equation
\eref{homentropicEulerEquationsa} and
\eref{homentropicEulerEquationsb} are the mathematical expressions
of conservation of mass and momentum respectively.  The homentropic
Euler equations makes the assumption that the entropy is constant
throughout the entire domain and thus the pressure can be expressed
solely as a function of the density \cite{LaneyCB:98a}.  This
differs from the isentropic Euler equations where the entropy is
constant along streamlines but not necessarily constant on the whole
domain.  It is found that the homentropic Euler equations are an
accurate predictor of gas dynamics behavior for low pressure.

Both equations have a nonlinear term that fits the general method
from section \ref{generalmethodsection}.  Apply
the method, break the nonlinear term apart, and then apply the
filter to the non-differentiated terms in the nonlinear terms to
obtain the equations
\begin{subequations}
\label{NMHE}
\begin{align}
\rho_t +\rhobar u_x +\ubar \rho_x=0\\
(\rho u)_t +(\overline{\rho u} )u_x +\ubar (\rho u)_x +P_x=0
\end{align}
\end{subequations}
with pressure defined as $P=\rho^\gamma$. These equations are now
referred to as the regularized homentropic Euler equations. For the following analysis, the only filter that is
considered is the Helmholtz filter \eref{helmholtzfilter}.

\subsection{Conservation of mass and
momentum}\label{NMHEconservationsection} The homentropic Euler
equations \eref{homentropicEulerEquations} are conservation laws for
mass and momentum. These equations are reflections of important
physical principles. In order to be physically relevant it is
desirable that any regularization of the homentropic Euler equations
would also preserve this structure. In section
\ref{generalconservationsection} we established that the technique
used will preserve the conservative structure of the original
equations.  Here we reverify that result for the regularized homentropic Euler equations
\eref{NMHE} by casting the equations into a conservative form using
the Helmholtz filter.

Consider the regularized homentropic Euler equations
\begin{subequations}
\label{NMHEfull}
\begin{align}
\label{NMHEfulla}\rho_t +\rhobar u_x +\ubar \rho_x=0\\
\label{NMHEfullb}(\rho u)_t +(\overline{\rho u}) u_x +\ubar (\rho
u)_x +P_x=0
\end{align}
\end{subequations}
with
\begin{subequations}
\begin{align}
\label{NMHEfullc}u=\ubar-\alpha^2 \ubar_{xx}\\
\label{NMHEfulld}\rho=\rhobar-\alpha^2 \rhobar_{xx}\\
\label{NMHEfulle}\rho u =\overline{\rho u}-\alpha^2 \overline{\rho
u}_{xx}.
\end{align}
\end{subequations}

By substituting in Equations \eref{NMHEfullc}, \eref{NMHEfulld}, and
\eref{NMHEfulle} and then regrouping the similar terms the regularized homentropic Euler equations can be rewritten as
\begin{subequations}
\label{NMHEconservativeform}
\begin{align}
\rho_t +\left(\rhobar \ubar -\alpha^2 (\ubar \rhobar_{xx}
+\rhobar \ubar_{xx})+\alpha^2 \ubar_x \rhobar_x \right)_x=0\\
(\rho u)_t +\left(\overline{\rho u} \ubar -\alpha^2 (\ubar
\overline{(\rho u)}_{xx} +\overline{\rho u} \ubar_{xx})+\alpha^2
\ubar_x \overline{\rho u}_x +P\right)_x=0.
\end{align}
\end{subequations}
This shows that the regularized homentropic Euler equations can be written in a conservative
form and preserve both mass $\int \rho$ and momentum $\int \rho u$
as well as the averaged mass $\int \bar{\rho}$ and averaged momentum
$\int \overline{\rho u}$. Examination of Equations
\eref{NMHEconservativeform} could lead to more geometric structure.
They may also lead to the application of numerical techniques
designed specifically for conservation laws.

\subsection{Traveling wave solution}\label{travelingwavesectionNMHE}
In this section, we establish that a previously known traveling wave
solution to the homentropic Euler equations is in fact a traveling
wave solution to the regularized homentropic Euler equations as well.  The traveling wave
solution that we examine for the homentropic Euler equations is a
single traveling shock.

First we establish notation used in this section.  The operator
$\left[ \cdot \right]$ will be used to quantify the difference
between the right and left limits of a function at a discontinuity.
For example if the limits of $u(x)$ at $x^*$ are defined as
\begin{eqnarray}
\lim_{x\to x^{*+}} u(x)=u_R\\
\lim_{x\to x^{*-}} u(x)=u_L,
\end{eqnarray}
then at $x^*$
\begin{equation}
\left[ u \right] = u_R-u_L.
\end{equation}

The Rankine-Hugoniot jump conditions establish that for weak
solutions of conservation laws of the form
\begin{equation}
\textbf{Q}_t+(\textbf{f})_x=0
\end{equation}
 a traveling discontinuity must have a speed $S$, where $S$ is
defined as \cite{LaneyCB:98a}
\begin{equation}\label{rankinehugoniot}
S=\frac{[\textbf{f}]}{[\textbf{Q}]}.
\end{equation}
Using this relationship it is easy to establish that the following
is a weak solution to the homentropic Euler equations
\eref{homentropicEulerEquations}
\begin{eqnarray}\label{travelingwavesolution}
u=\left\{\begin{array}{ll}
u_L & x<St \\
u_R & x\geq St
\end{array}\right.\\
\rho=\left\{\begin{array}{ll}
\rho_L & x<St \\
\rho_R & x\geq St
\end{array}\right.
\end{eqnarray}
The speed of the shock $S$ is twice defined, once for each
conservation law, by the Rankine-Hugoniot jump conditions as
\begin{equation}\label{RHmassconservationHE}
S=\frac{[\rho u]}{[\rho]}
\end{equation}
and
\begin{equation}\label{RHmomentumconservationHE}
S=\frac{[\rho u u +P]}{[\rho u]}.
\end{equation}
Thus $u_L, u_R, \rho_L,$ and $\rho_R$ must satisfy the relationship
\begin{equation}
\frac{[\rho u]}{[\rho]}=S=\frac{[\rho u u +P]}{[\rho u]}.
\end{equation}

Next we will establish that Equation \eref{travelingwavesolution} is
also a weak solution to the regularized homentropic Euler equations
\eref{NMHEconservativeform}. It is straightforward to establish that
with $u$ defined as in Equation \eref{travelingwavesolution} and
using the Helmholtz kernel definition Equation
\eref{helmholtzkernel} that $\ubar$ and $\rhobar$ are
\begin{eqnarray}\label{bartravelingwavesolution}
\ubar=\left\{\begin{array}{ll}
\frac{u_R-u_L}{2} \exp(\frac{x-St}{\alpha})+u_L & x<St \\
\frac{u_L-u_R}{2} \exp(-\frac{x-St}{\alpha})+u_R & x\geq St
\end{array}\right.\\
\rhobar=\left\{\begin{array}{ll}
\frac{\rho_R-\rho_L}{2} \exp(\frac{x-St}{\alpha})+\rho_L & x<St \\
\frac{\rho_L-\rho_R}{2} \exp(-\frac{x-St}{\alpha})+\rho_R & x\geq St
\end{array}\right.
\end{eqnarray}
Thus the values of $\ubar$ and $\rhobar$ at $x=St$ are
\begin{eqnarray}
\label{ubaraverage}
\ubar(St,t)=\frac{u_R+u_L}{2}\\
\label{rhobaraverage}
\rhobar(St,t)=\frac{\rho_R+\rho_L}{2}.
\end{eqnarray}

Additionally using the definition of the Helmholtz filter,
$u=\ubar-\alpha^2\ubar_{xx}$, we find that
\begin{equation}
\left[u\right]=\left[\ubar\right]-\alpha^2\left[\ubar_{xx}\right].
\end{equation}
Since $\ubar$ is continuous we establish that
\begin{equation}\label{ubarxxjump}
\left[\ubar_{xx}\right]=-\frac{1}{\alpha^2}\left[u\right],
\end{equation}
with the similar result
\begin{equation}\label{rhobarxxjump}
\left[\rhobar_{xx}\right]=-\frac{1}{\alpha^2}\left[\rho\right].
\end{equation}

Looking at the conservation form of the regularized homentropic Euler equations
\eref{NMHEconservativeform}, the Rankine-Hugoniot jump conditions
state that in order for Equation \eref{travelingwavesolution} to be
a weak solution, that the speed of the discontinuity, $S$, must
satisfy
\begin{equation}\label{RHmassconservationNMHE}
S=\frac{[\rhobar \ubar -\alpha^2 (\ubar \rhobar_{xx} +\rhobar
\ubar_{xx})+\alpha^2 \ubar_x \rhobar_x]}{[\rho]}
\end{equation}
and
\begin{equation}\label{RHmomentumconservationNMHE}
S=\frac{[\overline{\rho u} \ubar -\alpha^2 (\ubar \overline{(\rho
u)}_{xx} +\overline{\rho u} \ubar_{xx})+\alpha^2 \ubar_x
\overline{\rho u}_x +P]}{[\rho u]}.
\end{equation}
Since $\ubar, \rhobar, \ubar_x,$ and $\rhobar_x$ are continuous
Equation \eref{RHmassconservationNMHE} reduces to
\begin{equation}
S=\frac{ -\alpha^2 (\ubar [\rhobar_{xx}] +\rhobar
[\ubar_{xx}])}{[\rho]}.
\end{equation}
Substitute in Equations \eref{ubarxxjump} and \eref{rhobarxxjump} to
simplify further to
\begin{equation}
S=\frac{ (\ubar [\rho] +\rhobar [u]) }{[ \rho] }.
\end{equation}
Now, by using Equations \eref{ubaraverage} and \eref{rhobaraverage}
and the definition of $[\cdot]$ one can obtain
\begin{eqnarray}
S&=&\frac{\frac{1}{2} (u_R+u_L)(\rho_R-\rho_L ) +\frac{1}{2}
(\rho_R+\rho_L )(u_R-u_L)}{[\rho]}\\
&=&\frac{(\rho_R u_R-\rho_L u_L)}{[\rho]}\\
 &=&\frac{[\rho u]}{[\rho]},
\end{eqnarray}
which is identical to Equation \eref{RHmassconservationHE}.
Similarly Equation \eref{RHmomentumconservationNMHE} reduces to
Equation \eref{RHmomentumconservationHE}.  Thus for this specific
example, the Rankine-Hugoniot jump conditions are identical for the
homentropic Euler equations and the regularized homentropic Euler equations. This validates
our claim that Equation \eref{travelingwavesolution} is a traveling
weak solution for the regularized homentropic Euler equations.

\subsection{Shock Thickness}\label{shockthicknessectionNMHE}
With Equation \eref{travelingwavesolution} validated as a traveling
weak solution for the regularized homentropic Euler equations, we establish an analytical
result about shock thickness. Here the thickness of the shock is
defined to be the length over which 90\% of the amplitude change
takes place, centered at the point of inflection.

Again we note that from Equation \eref{travelingwavesolution},  and
the kernel definition of the Helmholtz filter
\eref{helmholtzkernel}, the traveling wave solution is
\begin{eqnarray}
\ubar=\left\{\begin{array}{ll}
\frac{u_R-u_L}{2} \exp(\frac{x-St}{\alpha})+u_L & x<St \\
\frac{u_L-u_R}{2} \exp(-\frac{x-St}{\alpha})+u_R & x\geq St
\end{array}\right.\\
\rhobar=\left\{\begin{array}{ll}
\frac{\rho_R-\rho_L}{2} \exp(\frac{x-St}{\alpha})+\rho_L & x<St \\
\frac{\rho_L-\rho_R}{2} \exp(-\frac{x-St}{\alpha})+\rho_R & x\geq St
\end{array}\right.
\end{eqnarray}

The thickness of the shock will then be $2\alpha b$, where b is the
value where
\begin{equation}
\int_{-b}^b \frac{1}{2}exp(-|x|)\, dx=\int_{-\alpha b}^{\alpha b}
\frac{1}{2\alpha}exp(\frac{-|x|}{\alpha}) \, dx=0.9.
\end{equation}
This length is independent of $\rho_R, \rho_L, u_R$ and $u_L$.  As
such, the thickness of the shock varies linearly on the parameter
$\alpha$.

\subsection{Diagonalization and eigenvalues}\label{eigenvaluesection}
The homentropic Euler equations have very clearly defined
eigenvalues, $u \pm a,$ where $a^2=\gamma \rho^{\gamma-1}$ is the
speed of sound.  This section examines how the proposed averaging
affects the eigenvalues of the system and thus the characteristic
speeds. In order to cast the equations in vector matrix form we
first rewrite the equations in their primitive variable form
\begin{subequations}
\label{primitivevariableform}
\begin{align}
\rho_t +\rhobar u_x +\ubar \rho_x=0\\
u_t+\ubar u_x +\underbrace{\left( \frac{\overline{\rho u} -
u\rhobar}{\rho} \right)}_{\beta} u_x + \frac{\gamma
\rho^{\gamma-1}}{\rho} \rho_x=0.
\end{align}
\end{subequations}
The equations can then be written in vector matrix form leading to
\begin{equation} \label{HEmatirxform}
\left[
  \begin{array}{c}
   \rho \\
    u \\
  \end{array}
\right]_t +\underbrace{\left[
   \begin{array}{cc}
     \ubar  & \rhobar \\
     \frac{a^2}{\rho} & \ubar +\beta \\
   \end{array}
 \right]}_A
 \left[
  \begin{array}{c}
    \rho \\
    u \\
  \end{array}
\right]_x =0.
\end{equation}
The eigenvalues of matrix $A$ are
\begin{equation}
\lambda^\pm = \ubar + \frac{\beta}{2} \pm \sqrt{\frac{\beta^2}{4}+a^2
\frac{\rhobar}{\rho}}.
\end{equation}
Examining quantities $\beta$ and $\frac{\rhobar}{\rho}$ it seems
apparent that as the filtering decreases $\beta \to 0$ and
$\frac{\rhobar}{\rho} \to 1$, thus regaining the original
eigenvalues.

The matrix $A$ can as be diagonalized and can thus be written in the form
\begin{equation}
A=Q \Lambda Q^{-1}
\end{equation}
where $\Lambda$ is a diagonal matrix with its diagonal entries being $\lambda^\pm$.
Using this we can write the regularized homentropic Euler equations in the characteristic form
\begin{equation}
\frac{\p \vb}{\p t}+ \Lambda \frac{\p \vb}{\p x}=0
\end{equation}
where
\begin{equation}
d \vb=Q^{-1} \left[\begin{array}{c}
   d \rho \\
   d u \\
  \end{array}\right].
\end{equation}
By diagonalizing $A$ we find that
\begin{equation}
Q^{-1}=\left[
{\begin{array}{rc}
1 & {\displaystyle \frac {\rho\,\beta + \sqrt{\rho\,(\rho\,\beta^{2} + 4\,a^{2}\,
\mathit{\rhobar})}}{2\,a^{2}}}  \\ [2ex]
1 & {\displaystyle \frac { \rho\,\beta - \sqrt{\rho\,(\rho\,\beta^{2} + 4\,a^{2}\,
\mathit{\rhobar})} }{2\,a^{2}}}
\end{array}}
\right].
\end{equation}
And thus we can define the quantities $v^\pm$ through the relation
\begin{equation}\label{vpmdefine}
d v^\pm=d \rho+  \frac{\rho\,\beta \pm \sqrt{\rho\,(\rho\,\beta^{2} + 4\,a^{2}\,\rhobar)}}{2\,a^{2}}\, du.
\end{equation}
We can then say that $d v^\pm=0$ along the characteristic $dx=\lambda^\pm \,dt.$

If $\alpha=0$, there would be no filtering and $\beta=0$.  In this case we would get
\begin{equation}
dv^\pm = d \rho \pm \frac{\rho}{a} d u=0 \quad \mbox{ along the characteristic } dx=u \pm a \,dt.
\end{equation}
which is the case for the homentropic Euler equations.  For the homentropic Euler equations the characteristic variables are able to be computed analytically and are
\begin{equation}
v^\pm=u\pm \frac{2a}{\gamma -1}.
\end{equation}

For non-zero $\alpha$'s there appears to be no straightforward way of integrating Equation \eref{vpmdefine}. Thus currently we have no analytical expression for the characteristic variables, though we have not spent much time investigation this possibility.

\subsection{Convergence to a weak solution}\label{weaksolutionsection}
It is the goal of our technique to develop new equations that
effectively capture the low wave mode behavior of the original
equations.  Thus it is desirable that the regularized homentropic Euler equations approximate
the homentropic Euler equations well.  Ideally we can show that as
the amount of filtering decreases we regain the original equations.
One crucial step in determining this is proving that as $\alpha \to
0$ the solutions to the regularized homentropic Euler equations converge to a weak solution
of the homentropic Euler equations. This, among other things, proves
that as $\alpha \to 0$, the shocks produced by the regularized homentropic Euler equations
travel at the same speed as those of the homentropic Euler
equations, which is desirable.

As of yet, we have not been able to extend our existence theorems of
the regularized Burgers equations \eref{CFB} to the regularized homentropic Euler equations
\eref{NMHE}. Without this theorem we are forced to make several
assumptions that we consider modest, considering the numerical
evidence presented  in section \ref{numericalresultssection}. We
assume that for every $\alpha > 0$ there exists a solution to
Equations \eref{NMHEfull}. Beyond that it is assumed that a
subsequence of those solutions converge in $L^1_{loc}$ and the
solutions are bounded independent of $\alpha.$ The following
summarizes these assumptions.
\begin{subequations}\label{NMHEassumptions}
\begin{align}
||u||_\infty < U\\
||\rho||_\infty < R\\
\lim_{\alpha \to 0}  u = \tilde{u} \mbox{ in }L^1_{loc}\\
\lim_{\alpha \to 0} \rho = \tilde{\rho} \mbox{ in }L^1_{loc}
\end{align}
\end{subequations}

With these assumptions we are able to prove that the solutions to
the regularized homentropic Euler equations \eref{NMHE} will converge to weak solutions of
the homentropic Euler equations \eref{homentropicEulerEquations}.
The examination of the claim is done with the Helmholtz filter, with
$\ubar$ defined as
\begin{equation}
\ubar=u \ast g^\alpha
\end{equation}
with
$$g(x)=\frac{1}{2}exp(-|x|)$$
and thus
$$g^\alpha(x)=\frac{1}{2\alpha}exp(-\frac{|x|}{\alpha}).$$
The unfiltered and filtered velocities can also be related by
\begin{equation}
u=\ubar-\alpha^2 \ubar_{xx}.
\end{equation}
Both definitions are used.  The following bounds are easily
established by examining the kernel.
$$||g||_1 =1$$
and
$$\left|\left|\frac{\p}{\p x} g\right|\right|_1 =1,$$
and thus
$$\left|\left|\frac{\p}{\p x} g^{\alpha}\right|\right|_1 =\frac{1}{\alpha}.$$
These bounds combined with Young's inequality \cite{HunterJK:01a}
and the assumptions \eref{NMHEassumptions} give the following
estimates
\begin{align}
||\ubar||_\infty < U\\
||\rhobar||_\infty < R\\
||\mbar||_\infty < UR\\
||\ubar_x||_\infty < \frac{1}{\alpha} U\\
||\rhobar_x||_\infty< \frac{1}{\alpha} R\\
||\mbar_x||_\infty < \frac{1}{\alpha} UR
\end{align}
Along with these estimates, the last piece needed is taken from
Duoandikoetxea \cite{DuoandikoetxeaJ:99a}.  The following lemma is a
restatement of Duoandikoetxea's Theorem 2.1 from page 25.
\begin{lemma}
\label{HEconvolutionlemma} Let $g$ be an integrable function on
$\mathbb{R}$ such that $\int g =1$.  Define
$g^\alpha=\frac{1}{\alpha} g(\frac{x}{\alpha})$.  Then
$$\lim_{\alpha \to 0}  ||g^\alpha \ast f - f||_p = 0$$  if $f \in L^p, 1\leq p < \infty$ and uniformly (i.e. when $p=\infty$) if $f \in C_0(\mathbb{R})$.
\end{lemma}
With lemma \ref{HEconvolutionlemma} the convergence to weak
solutions can now be proven.  Begin by multiplying Equations
\eref{NMHEconservativeform} by a test function $\phi$ and integrate
over time and space.  It is assumed that $\phi$ has an infinite
number of bounded and continuous derivatives and is compactly
supported. Doing this we obtain the equations
\begin{subequations}
\begin{align}
\int_\mathbb{R} \int_0^T \rho_t \phi +\left(\rhobar \ubar -\alpha^2
(\ubar \rhobar_{xx}
+\rhobar \ubar_{xx})+\alpha^2 \ubar_x \rhobar_x \right)_x \phi \, dt \, dx=0\\
\int_\mathbb{R} \int_0^T (\rho u)_t \phi +\left(\mbar \ubar
-\alpha^2 (\ubar \overline{(\rho u)}_{xx} +\mbar
\ubar_{xx})+\alpha^2 \ubar_x \mbar_x \right) +P)_x \phi \, dt \,
dx=0.
\end{align}
\end{subequations}
Integrate by parts to obtain
\begin{subequations}\label{beforelimitequations}
\begin{align}
\int_\mathbb{R} \int_0^T \rho \phi_t +\left(\rhobar \ubar \right) \phi_x \, dt \, dx=\int_\mathbb{R} \int_0^T \left(  \alpha^2(\ubar \rhobar_{xx} +\rhobar \ubar_{xx})- \alpha^2\ubar_x \rhobar_x \right) \phi_x \, dt \, dx\\
\int_\mathbb{R} \int_0^T (\rho u) \phi_t  +\left(\mbar \ubar +P
\right) \phi_x \, dt \, dx = \int_\mathbb{R} \int_0^T   \left(
\alpha^2(\ubar \overline{(\rho u)}_{xx} +\mbar \ubar_{xx})- \alpha^2
\ubar_x \mbar_x \right) ) \phi_x \, dt \, dx.
\end{align}
\end{subequations}
Clearly if the right  hand side of Equations
(\ref{beforelimitequations}a) and (\ref{beforelimitequations}b)
limit to zero then the limit of $\rho$ and $u$ is a weak solution to
the homentropic Euler equations.  Begin by examining the term
\begin{equation}
\int_\mathbb{R} \int_0^T  \alpha^2 \ubar \rhobar_{xx} \phi_x \, dt
\, dx
\end{equation}
Substitute the definition of the averaging so that
 $ \alpha^2 \rhobar_{xx} = \rhobar-\rho$ and examine the
quantity
 \begin{equation}
\int_\mathbb{R} \int_0^T  \ubar (\rhobar-\rho) \phi_x \, dt \, dx.
\end{equation}
It was established that $\ubar$ and $\phi_x$ are bounded and by
lemma \ref{HEconvolutionlemma} $(\rhobar-\rho)$  converges in $L^1$
to zero. Thus we find
\begin{equation}
\lim_{\alpha \to 0} \int_\mathbb{R} \int_0^T  \alpha^2 \ubar
\rhobar_{xx} \phi_x \, dt \, dx=\int_\mathbb{R} \int_0^T  \ubar
(\rhobar-\rho) \phi_x \, dt \, dx=0,
\end{equation}
and the similar terms can be treated likewise.

Next examine the term
\begin{equation}
\int_\mathbb{R} \int_0^T \alpha^2 \ubar_x \mbar_x   \phi_x \, dt \,
dx.
\end{equation}
Perform an integration by parts to obtain
\begin{equation}\label{equationwithoutconvientname1}
-\int_\mathbb{R} \int_0^T \alpha^2 \ubar \overline{(\rho u)}_{xx}
\phi_x \, dt \, dx - \int_\mathbb{R} \int_0^T \alpha^2 \ubar \mbar_x
\phi_{xx} \, dt \, dx.
\end{equation}
The first term in Equation \eref{equationwithoutconvientname1}
limits to zero from the steps shown above.  The second term can be
bounded with the estimates found from Young's inequality.
\begin{align}
\left| \left| \int_\mathbb{R} \int_0^T \alpha^2 \ubar \mbar_x   \phi_{xx} \, dt \, dx \right| \right| &\leq \int_\mathbb{R} \int_0^T   \alpha^2 U \frac{1}{\alpha}  UR  | \phi _{xx}| \, dt \, dx\\
&=\alpha U^2 R ||\phi _{xx}||_1
\end{align}
and thus limits to zero. The similar terms can be treated likewise.
 Thus we see that solutions to Equations \eref{NMHEconservativeform}
converge to weak solutions of the homentropic Euler equations as
$\alpha \to 0$.

\subsection{Numerics}\label{numericssection}In this section, we discuss the numerical techniques we used to simulate the regularized homentropic Euler equations and the homentropic Euler equations. Since the regularized homentropic Euler equations appear to be regularized and its solutions are smooth, we are able to use a pseudo-spectral method. The homentropic Euler equations tend to form shocks, so we use well-established techniques to handle this behavior.

\subsubsection{Numerical simulations of the regularized homentropic Euler equations}
For the regularized homentropic Euler equations, much like in turbulence simulation, we wish only to resolved the
averaged quantities in our numerical simulations.  To obtain the
equations with only the smooth variables, we first apply the
Helmholtz filter to Equations \eref{NMHE} and add $(\rhobar
\ubar)_x$ or $(\mbar \ubar)_x$ to both sides.
\begin{subequations}
\begin{align}
\rhobar_t +  \left(\rhobar \ubar               \right)_x      = \left(\rhobar \ubar\right)_x - \overline{(\rhobar u_x +\ubar \rho_x)}\\
\mbar_t   +  \left(\mbar   \ubar     +\bar{P}   \right)_x =
\left(\mbar \ubar\right)_x - \overline{(\mbar u_x +\ubar (\rho
u)_x)}.
\end{align}
\end{subequations}
Then use the definition of the Helmholtz filter
\eref{helmholtzfilter} to manipulate the equations to
\begin{subequations}
\begin{align}
\rhobar_t +  \left(\rhobar \ubar               \right)_x      = \overline{(1-\alpha^2\p_x^2)\left(\rhobar \ubar\right)_x - \rhobar (\ubar_x-\alpha^2 \ubar_{xxx}) -\ubar (\rhobar_x-\alpha^2 \rhobar_{xxx})}\\
\mbar_t   +  \left(\mbar   \ubar     +\bar{P}   \right)_x =
\overline{(1-\alpha^2\p_x^2) \left(\mbar \ubar\right)_x - \mbar
(\ubar_x-\alpha^2 \ubar_{xxx}) -\ubar ((\mbar)_x-\alpha^2
(\mbar)_{xxx})}.
\end{align}
\end{subequations}
One can then simplify to obtain
\begin{subequations}
\label{NMHEsmoothonlyform}
\begin{align}
\rhobar_t +  \left(\rhobar \ubar               \right)_x      = -3\alpha^2 \overline{(\ubar_x \rhobar_x)_x}\\
\mbar_t   +  \left(\mbar   \ubar     +\bar{P}   \right)_x      =
-3\alpha^2 \overline{(\ubar_x \mbar_x)_x}.
\end{align}
\end{subequations}
The equations are now fully in terms of the averaged quantities with
the right hand side being the regularizing terms.

In order to reduce the potential for numerical dissipation we use a
pseudo-spectral method to solve Equations \eref{NMHEsmoothonlyform}.
We advance the equations in time with an explicit,
Runge-Kutta-Fehlberg predictor/corrector (RK45) \cite{Holm:03a}. The
initial time step is chosen small enough to achieve stability, and
is then varied by the code using the formula
\begin{equation}
h_{i+1}=\gamma h_{i} \left( \frac{\varepsilon h_{i}} {||\bar{\rho}_i
-\hat{\rho}_i||_{2} }\right)^\frac{1}{4}.
\end{equation}
Thus the new time step is chosen from the previous time step and the
amount of error between the predicted density, $\bar{\rho}$ and the
corrected density $\hat{\rho}$. The relative error tolerance was
chosen at $\varepsilon=10^{-4}$ and the safety factor $\gamma=0.9$.
If the new time step chosen was found to violate the CFL condition
the time step was chosen according to the velocity speed and the
speed of sound with CFL number $0.5$.

Spatial derivatives and the inversion of the Helmholtz operator were
computed in the Fourier domain.  The terms were converted into the
Fourier domain using a Fast Fourier Transform, multiplied by the
appropriate term and then converted back into the physical domain.

\begin{figure}[!ht]
\begin{center}
  \includegraphics[width=.5\linewidth]{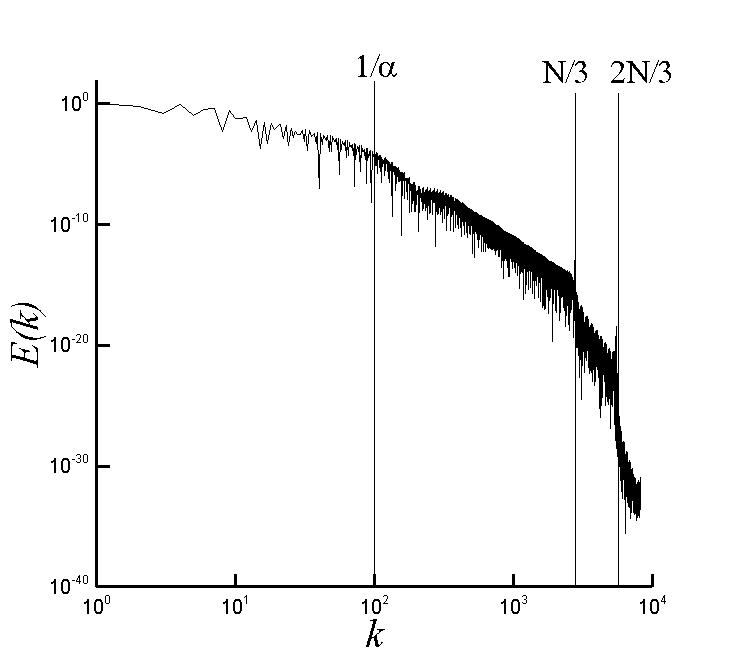}
\caption{The spectral energy of $u$.  The vertical lines represent
three significant wave numbers.  From left to right the lines
represent $\frac{1}{\alpha}$, $\frac{N}{3}$, and $\frac{2N}{3}$,
where $N$ is the highest Fourier wave mode simulated. The wave mode
$\frac{N}{3}$ is the highest wave mode that is ever zeroed out in
the simulations. Here with $\alpha =0.01$ you can see that zeroing
of wave modes takes place an order of magnitude higher than where
the filtering has its effect. This graph was taken from simulations
of examples problem \ref{NMHEExample}. Here $t=0.1$, $N=2^{13}$, and
$n=2^{14}$, where $N$ is the number of Fourier modes resolved and
$n$ is the number of grid points.} \label{NMHEspectralfigure}
\end{center}
\end{figure}

A number of high resolution simulations were done at the resolution
of $2^{14}=65536$ grid points on the domain $[0, 2\pi]$.  We found
that some long term simulations may suffer from long term numerical instabilities.  After approximately 10,000 time steps, we would notice energy
building up in the high wave modes near the resolution boundary. If not checked, these would contaminate the results of lower
wave modes. In order to control this long term instability, every
200 time steps we zero out all the wave modes with wave number
higher that $\frac{N}{3}$ where $N$ is the highest Fourier wave mode
simulated. Wave modes higher than $\frac{2N}{3}$, are set to zero at
every nonlinear multiplication to prevent aliasing.  Numerical runs
were conducted for values $\alpha=0.10, 0.09, ..., 0.02, 0.01.$ In
the worst case scenario, $\alpha=0.01,$ the wave modes that are
zeroed are over an order of magnitude higher than
$\frac{1}{\alpha}$, as seen in figure \ref{NMHEspectralfigure}.

We attribute this error to the right hand side of Equations
\eref{NMHEsmoothonlyform} for the following reasons. Examine the
term
\begin{equation}
-3\alpha^2 \overline{(\ubar_x \rhobar_x)_x}.
\end{equation}
Expand the term to
\begin{equation}
-3\alpha^2 \overline{\ubar_{xx} \rhobar_x +\ubar_x \rhobar_{xx}}.
\end{equation}
and specifically notice the $\ubar_x \rhobar_{xx}$ term.  This term
can be considered as a viscosity-like term that  changes signs with
$\ubar_x$.  Now along the expansion wave $\ubar_x>0$ and thus there
is a negative second derivative or viscosity-like term. Negative
viscosity is inherently unstable, so we attribute the numerical
instabilities to this.  This is similar to the backscatter of energy to larger scales that is observed in turbulence.

\subsubsection{Numerics for the homentropic Euler
equations}\label{HEnumerics}
For the numerical simulations of the
homentropic Euler equations there are many established, available techniques\cite{LaneyCB:98a}. We chose to use the Richtmyer method, a well-established if low order method, was utilized as described by \cite{LaneyCB:98a}. This method is a second-order, finite-difference
scheme and employs an artificial viscosity. This
method requires an artificial viscosity for stability when examining
the Riemann problem.  Several different values of $\nu$ were tested
to see that the value did not significantly affect the solutions on
the time interval examined. For the numerical simulations shown
here, the artificial viscosity was set at $\nu=0.08$. For reference, the simulations were done with  $2^{14}$ grid points on a $[0, 2 \pi]$ domain.

\subsection{Numerical results}\label{numericalresultssection}
This section examines some numerical simulations performed on
Equations \eref{NMHEsmoothonlyform} with the technique described in
the previous section. A shock tube or Riemann problem exhibits both
shocks and expansion waves, some of the key behaviors of the
homentropic Euler equations. Our pseudo-spectral method enforces
periodic boundary conditions, so we created two pressure jumps in
the initial conditions resolving this issue. Thus essentially a
double shock tube problem is considered where there are two
discontinuities in the initial conditions to make the left and right
side boundary conditions identical. The example problem considered
is
\begin{eqnarray}
u_0(x)=&0\\
 a_0(x)=&\left\{\begin{array}{ll}
1 & 0<x\leq\frac{2\pi}{3} \\
2 &  \frac{2\pi}{3}<x\leq\frac{4\pi}{3} \\
1 & \frac{4\pi}{3}<x\leq 2\pi
\end{array}\right.
\end{eqnarray}
which for $\gamma =1.4$, the constant for air, is equivalent to
\begin{eqnarray}\label{NMHEExample}
u_0(x)=&0\\
 \rho_0(x)=&\left\{\begin{array}{ll}
0.43120 & 0<x\leq\frac{2\pi}{3} \\
13.7984 &  \frac{2\pi}{3}<x\leq\frac{4\pi}{3} \\
0.43120 & \frac{4\pi}{3}<x\leq 2\pi
\end{array}\right. .
\end{eqnarray}

In our previous work, we established that for the CFB equations
\eref{CFB} initial conditions with discontinuities were excluded in order to avoid non-entropic behavior \cite{NorgardG:08a}. By regularizing the initial conditions it was proven that the solution to the CFB equations would be regularized for all time. This was done by averaging the initial conditions with the same filter used on the velocity. We use this same approach here. Thus when the filter is applied to the entire equations to
obtain Equations \eref{NMHEsmoothonlyform}, the initial conditions
are filtered twice with the same filter used on the equations.

\begin{figure}[!ht]
\begin{center}
\begin{minipage}{0.48\linewidth} \begin{center}
  \includegraphics[width=.9\linewidth]{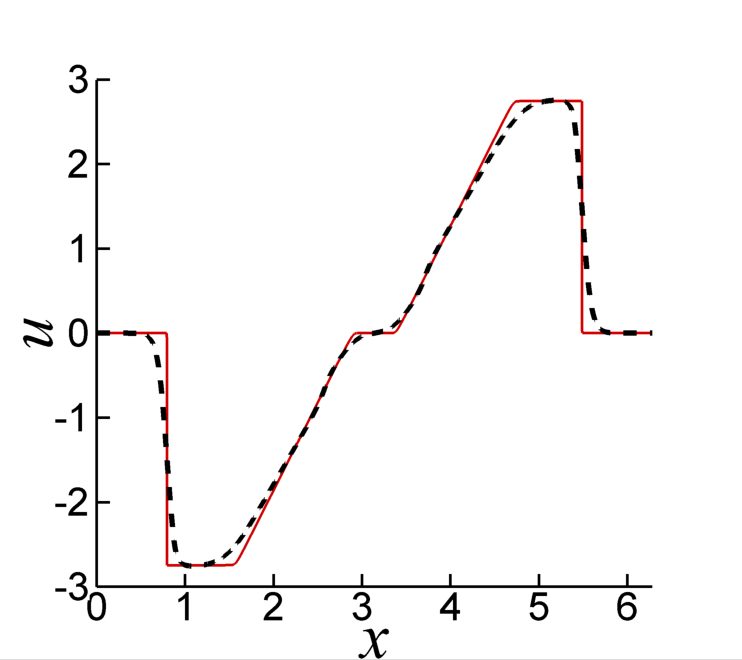}
\end{center} \end{minipage}
\begin{minipage}{0.48\linewidth} \begin{center}
  \includegraphics[width=.9\linewidth]{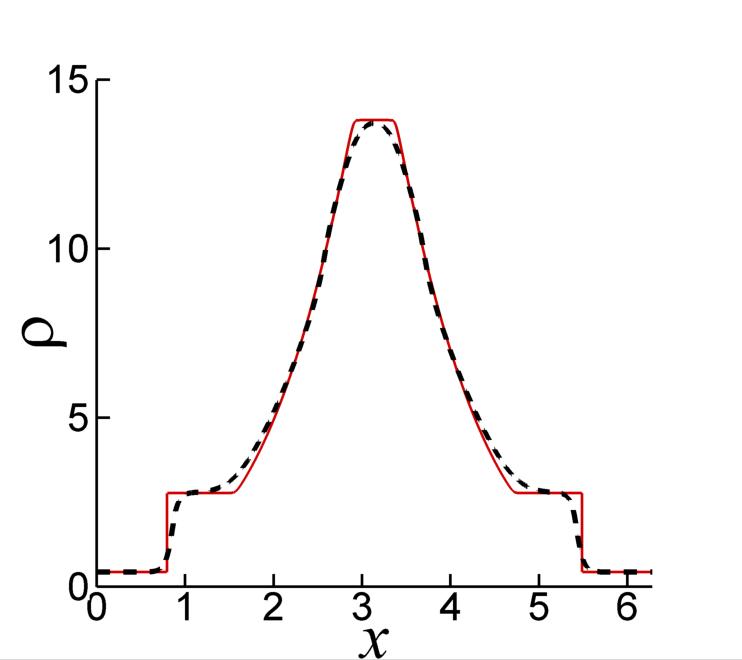}
\end{center} \end{minipage}\\
\begin{minipage}{0.48\linewidth}\begin{center} (a) \end{center} \end{minipage}
\begin{minipage}{0.48\linewidth}\begin{center} (b) \end{center}
\end{minipage}
\caption{This figure shows a numerical simulation of the regularized homentropic Euler equations (dashed line) plotted against the solution to the
homentropic Euler equations (solid line).  Here the value of
$\alpha=0.05$.  In both figures, it is clear that the regularized homentropic Euler equations
are capturing both the expansion wave and shock behavior.  (a) The
velocity. (b) The density.} \label{doubleriemanna05}
\end{center}
\end{figure}

\begin{figure}[!ht]
\begin{center}
\begin{minipage}{0.48\linewidth} \begin{center}
  \includegraphics[width=.9\linewidth]{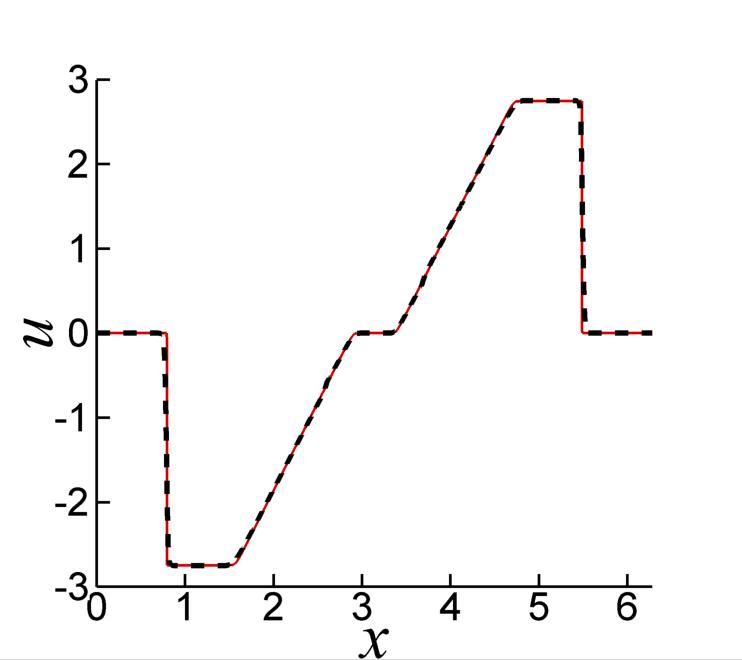}
\end{center} \end{minipage}
\begin{minipage}{0.48\linewidth} \begin{center}
  \includegraphics[width=.9\linewidth]{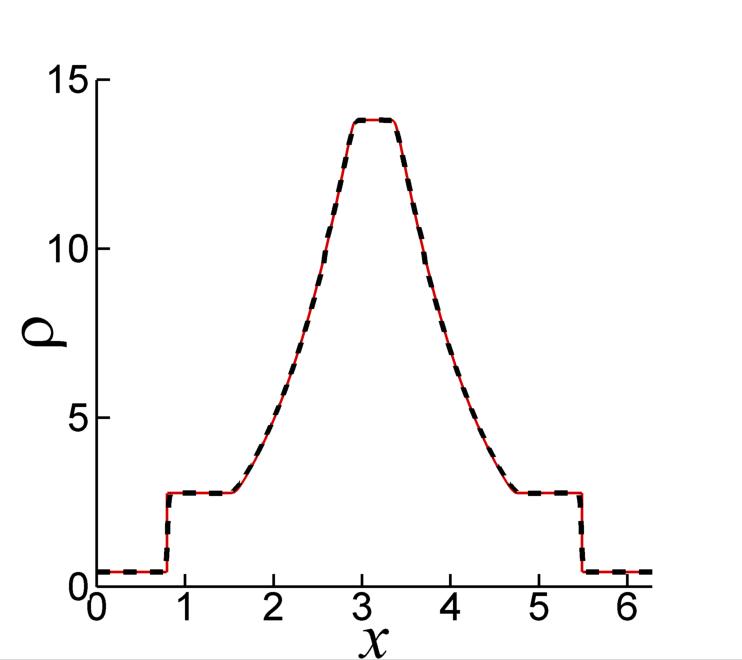}
\end{center} \end{minipage}\\
\begin{minipage}{0.48\linewidth}\begin{center} (a) \end{center} \end{minipage}
\begin{minipage}{0.48\linewidth}\begin{center} (b) \end{center}
\end{minipage}
\caption{This figure shows a numerical simulation of the regularized homentropic Euler equations (dashed line) plotted against the solution to the
homentropic Euler equations (solid line).  Here the value of
$\alpha=0.01$. In both figures, it is clear that the regularized homentropic Euler equations
are capturing both the expansion wave and shock behavior.  With the
lower value of $\alpha$ the fit is much closer. (a) The velocity.
(b) The density.} \label{doubleriemanna01}
\end{center}
\end{figure}

In the simulation of the double shock tube problem the regularized homentropic Euler equations are displaying similar behavior to the homentropic Euler
equations.  Figures \ref{doubleriemanna05} and
\ref{doubleriemanna01} show the solutions to the two sets of
equations imposed on each other.  One can see that the solutions to
the regularized homentropic Euler equations capture both the expansion wave and the shock
front.  The solutions to the regularized homentropic Euler equations are seen to be smooth
with the solutions tightening to the solutions of the homentropic
Euler equations as $\alpha$ decreases.

\begin{figure}[!ht]
\begin{center}
  \includegraphics[width=.5\linewidth]{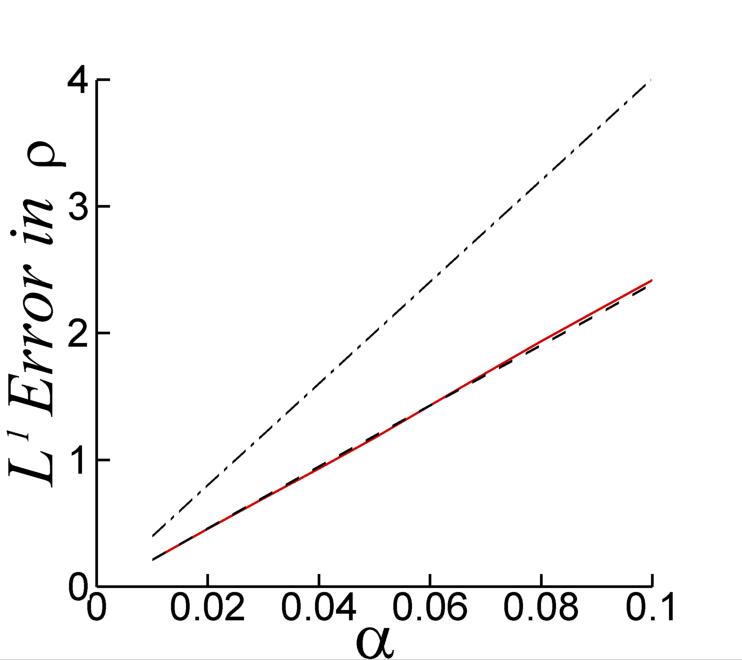}
\caption{This figure displays the difference between the density in
solutions of the regularized homentropic Euler equations and the solution of the homentropic
Euler equations in the $L^1$ norm as $\alpha \to 0$.  As $\alpha \to
0$ the difference in the solutions also approaches zero.  The
measurements were taken for $\alpha=0.01, 0.02, ..., 0.1$ at times
$t=0$, \dashdot, $t=0.2$ \solid, and $t=0.4$ \dashed.}
\label{NMHEl1rhoerror}
\end{center}
\end{figure}

\begin{figure}[!ht]
\begin{center}
  \includegraphics[width=.5\linewidth]{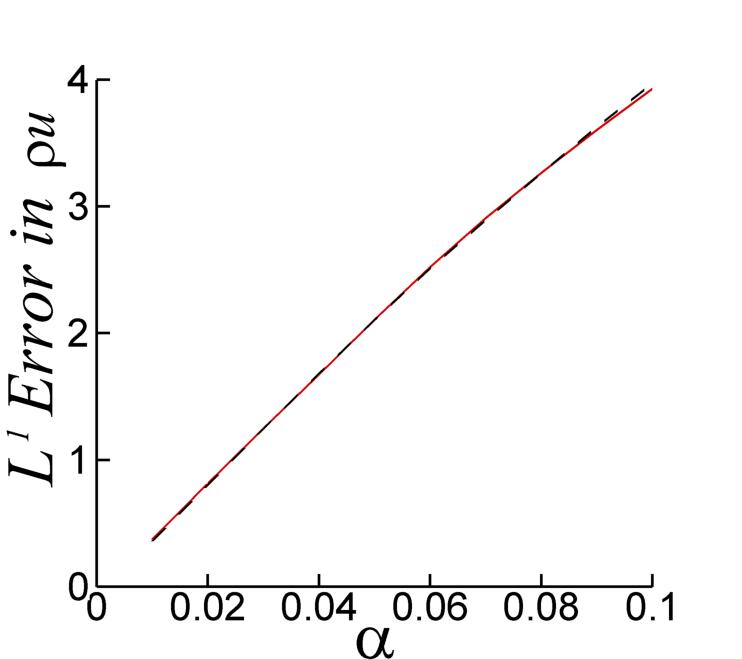}
\caption{This figure displays the difference between the momentum in
solutions of the regularized homentropic Euler equations and the solution of the homentropic
Euler equations in the $L^1$ norm as $\alpha \to 0$.  As $\alpha \to
0$ the difference in the solutions also approaches zero.  The
measurements were taken for $\alpha=0.01, 0.02, ..., 0.1$ at times
 $t=0.2$ \solid, and $t=0.4$ \dashed.}
\label{NMHEl1merror}
\end{center}
\end{figure}

Now we check the convergence of the solutions of the regularized homentropic Euler equations
to the solution of the homentropic Euler equations as $\alpha \to
0$.  Figures \ref{NMHEl1rhoerror} and \ref{NMHEl1merror} show that
as $\alpha \to 0$ the error in the $L^1$ norm appears to be
approaching zero for the example problem. This  suggests that the
solutions of the regularized homentropic Euler equations converge to the solutions of the
homentropic Euler equations.

\subsection{Kinetic energy rates}\label{kineticenergysectionNMHE}
In addition to checking solution profiles, we examine the effect that our regularization technique has upon the kinetic energy. For the homentropic Euler equations we define kinetic energy as $\frac{1}{2} \rho u^2.$  For the regularized homentropic Euler equations there are three different averaged quantities, $\rhobar$, $\ubar$, and $\mbar$.   With these quantities, kinetic energy can be defined in a variety of ways.   In this section we examine a kinetic energy  with unfiltered terms, $\frac{1}{2} \rho u^2$, and a kinetic energy with filtered terms, $\frac{1}{2} \rhobar \ubar^2$.  We have also examined the possible kinetic energies $\frac{1}{2} \mbar \ubar$ and $\frac{\mbar^2}{2 \rhobar}$, but for the example we are considering the difference between them and $\frac{1}{2} \rhobar \ubar^2$ was of the order $10^{-14}$ and thus negligible.  For flows with more small scale behavior, we would expect this difference to be more significant.

\begin{figure}[!ht]
\begin{center}
\begin{minipage}{0.48\linewidth} \begin{center}
  \includegraphics[width=.9\linewidth]{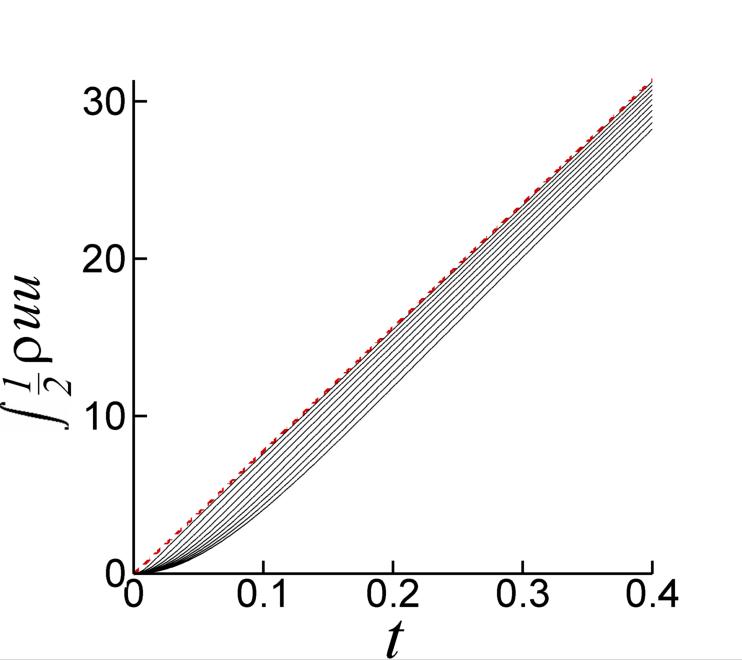}
\end{center} \end{minipage}
\begin{minipage}{0.48\linewidth} \begin{center}
  \includegraphics[width=.9\linewidth]{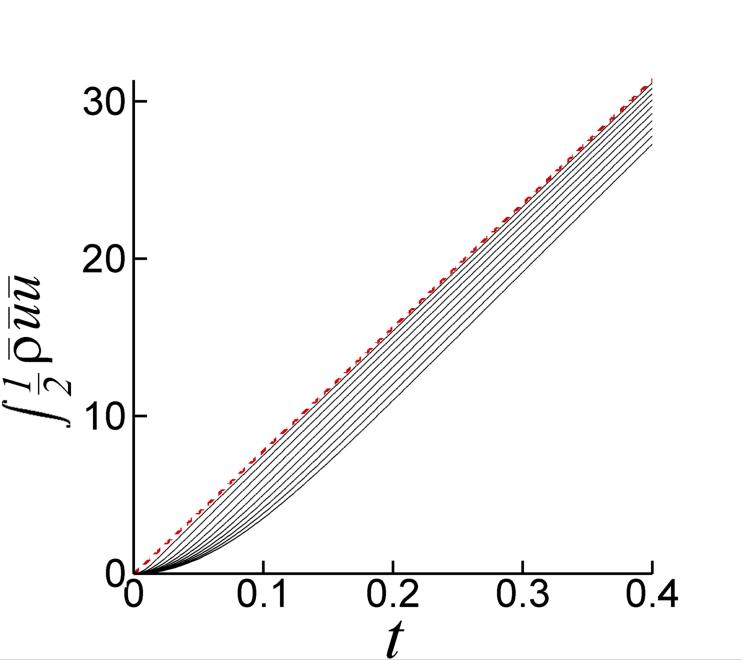}
\end{center} \end{minipage}\\
\begin{minipage}{0.48\linewidth}\begin{center} (a) \end{center} \end{minipage}
\begin{minipage}{0.48\linewidth}\begin{center} (b) \end{center}
\end{minipage}
\caption{The kinetic energy of the homentropic Euler equations and the regularized homentropic Euler equations. The energy for the true solution of the homentropic Euler equations is shown as a dashed red line.  The simulation of the regularized homentropic Euler equations for $\alpha=0.01, 0.02, ..., 0.1$ are shown as solid black lines.  The bottommost line represents $\alpha=0.1$.  As $\alpha$ decreases the energy approaches the energy of the homentropic Euler equations.  (a) These are plots of the unfiltered kinetic energies, $\frac{1}{2} \rho u^2$.  (b)  These are plots of the filtered kinetic energies, $\frac{1}{2} \rhobar \ubar^2$.  When examined the plots of $\frac{1}{2} \mbar \ubar$ and $\frac{\mbar^2}{2 \rhobar}$ were identical to this one. } \label{kinecticenergyfigure}
\end{center}
\end{figure}

For the shock tube problem, the kinetic energy of the system clearly starts at zero.  For the homentropic Euler equations the solution is self similar, depending only on the variable $\frac{x}{t}$.  Thus the kinetic energy for the homentropic Euler equations will be a linear function of time.  This can be seen in figure \ref{kinecticenergyfigure}.  We find that the energies of the regularized homentropic Euler equations mimic that of the homentropic Euler equations.  In the simulations of the regularized homentropic Euler equations there is a brief period where the energy growth is curved before it appears to behave linearly.  We attribute this to the averaging of the initial conditions.  As we would expect, as $\alpha$ decreases the energies grow closer to those of the homentropic Euler equations.

\section{The one-dimensional Euler equations}\label{NMEsection}
While the homentropic Euler equations are good for low pressures, to
have real impact we want our technique to be able to capture the
behavior of the full one-dimensional Euler equations.  With the regularized homentropic Euler equations showing promise, we attempt to use the same general
methodology on the one-dimensional Euler equations. The
one-dimensional Euler equations consist of three conservation laws
paired with a constitutive law. The conservation laws are
conservation of mass, momentum, and energy. For the constitutive
law, this paper is considering an ideal gas. Thus the
one-dimensional Euler equations considered, in conservation form,
are
\begin{subequations}
\label{EulerEquations}
\begin{align}
\label{EulerEquationsa} \rho_t+(\rho u)_x=0\\
\label{EulerEquationsb} (\rho u)_t+ (\rho u u+P)_x=0\\
\label{EulerEquationsc} (\rho e)_t+ (\rho e u+uP)_x=0\\
\label{EulerEquationsd} P=(\gamma -1) \left(\rho e -\frac{1}{2} \rho
u^2\right)
\end{align}
\end{subequations}

The technique described in section \ref{generalmethodsection}
applies the product rule to nonlinear terms and then the
non-differentiated quantities are spatially filtered. Again this is
to reduce the production of higher wave modes as time progresses,
although as in the regularized homentropic Euler equations, we will see that an artificial
viscosity is needed for numerical stability. When our averaging
technique is applied, the new equations are
\begin{subequations}
\label{NME}
\begin{align}
\label{NMEa} \rho_t+\rhobar u_x+ \ubar \rho_x=0\\
\label{NMEb} (\rho u)_t+ \mbar u_x+ \ubar (\rho u)_x+P_x=0\\
\label{NMEc} (\rho e)_t+ \overline{\rho e} u_x+ \ubar (\rho e)_x +\overline{P} u_x+ \ubar P_x=0\\
\label{NMEd} P=(\gamma -1) \left(\rho e -\frac{1}{2} \rho u^2\right)
\end{align}
\end{subequations}
which we refer to as regularized Euler equations. The
general method established in section \ref{generalmethodsection}
does not address exactly how to handle the pressure terms.  In the
conservation of momentum equation \eref{EulerEquationsb} the
pressure term is left unaffected as in the regularized homentropic Euler equations.  However,
the $(uP)_x$ term in the conservation of energy equation
\eref{EulerEquationsc} is averaged using the general method. We have
found in earlier numerical simulations that not performing the
averaging technique on this term lead to some nonphysical behavior.

The rest of this paper is dedicated to the examination of the regularized Euler equations. The next section shows that the regularized Euler equations conserve
mass, momentum, and energy, by casting the equations into a
conservative form using the Helmholtz filter. Section
\ref{travelingwavesectionNME} establishes a traveling wave solution
using the same techniques as before.  Section
\ref{NMEweaksolutionsection} proves that with modest assumptions the
regularized Euler  equations converge to a weak solution of the Euler equations as
$\alpha \to 0$.  Finally sections \ref{NMEnumericssection} and
\ref{NMEnumericalresultssection} examine some numerical simulations
and their results.

\subsection{Conservation of mass and momentum}\label{NMEconservationsection}
The Euler equations \eref{EulerEquations} are conservation laws that
address conservation of mass, momentum, and energy.  This section
show that the regularized Euler equations preserve these same quantities by
casting the equations into a conservative form. It would be
sufficient to refer the reader back to sections
\ref{generalconservationsection} and \ref{NMHEconservationsection}
and note that our method does not disturb the conservation structure
of the Euler equations. However, explicitly showing this for the regularized Euler equations for the Helmholtz filter has merit as it demonstrates a
new form of the equations which are of interest.

To review, the Helmholtz filter the non-filtered quantities can be
expressed by their filtered counterparts by Equation \eref{helmholtzfilter}.
Using this filter, Equations \eref{NME} can be rewritten as
\begin{subequations}
\label{NMEconservationform}
\begin{align}
\rho_t +&\left[\rhobar \ubar -\alpha^2 (\ubar \rhobar_{xx}+\rhobar \ubar_{xx})+\alpha^2 \ubar_x \rhobar_x \right]_x=0\\
(\rho u)_t +&\left[\overline{\rho u} \ubar -\alpha^2 (\ubar \overline{(\rho u)}_{xx} +\overline{\rho u} \ubar_{xx})+\alpha^2 \ubar_x \overline{\rho u}_x +P\right]_x=0\\
(\rho e)_t +&\left[\overline{\rho e} \ubar -\alpha^2 (\ubar \overline{(\rho e)}_{xx} +\overline{\rho e} \ubar_{xx})+\alpha^2 \ubar_x \overline{\rho e}_x + \right. \nonumber \\
                  +&\left.\overline{P} \ubar -\alpha^2 (\ubar \overline{P}_{xx} +\overline{P} \ubar_{xx})+\alpha^2 \ubar_x \overline{P}_x \right]_x=0
\end{align}
\end{subequations}
In this conservation form it is easy to see that mass $\int \rho$,
momentum $\int \rho u$, and energy $\int \rho e$ are conserved.

\subsection{Traveling wave solution}\label{travelingwavesectionNME}
In section \ref{travelingwavesectionNMHE} it was established that
certain traveling weak solutions to the homentropic Euler equations
were also solutions to the regularized homentropic Euler equations.  The section establishes
a similar result for the Euler and regularized Euler equations using the same
techniques. Again the operator $\left[ \cdot \right]$ will be used
to quantify the difference between the right and left limits of a
function at a discontinuity so that
\begin{equation}
\left[ u \right] = u_R-u_L.
\end{equation}

Again we examine a traveling shock of the form
\begin{eqnarray}\label{travelingwavesolutionEuler}
\rho=\left\{\begin{array}{ll}
\rho_L & x<St \\
\rho_R & x\geq St
\end{array}\right. \\
\rho u=\left\{\begin{array}{ll}
M_L & x<St \\
M_R & x\geq St
\end{array}\right. \\
\rho e=\left\{\begin{array}{ll}
E_L & x<St \\
E_R & x\geq St
\end{array}\right. .
\end{eqnarray}
Using the Rankine-Hugoniot jump conditions \eref{rankinehugoniot} we
can establish that at a discontinuity a weak solution to the Euler
equations \eref{EulerEquations} must satisfy
 \begin{subequations}
\label{RHfulleuler}
\begin{eqnarray}
\label{RHmassconservationE}
S&=&\frac{[\rho u]}{[\rho]}\\
 \label{RHmomentumconservationE}
S&=&\frac{[\rho u u +P]}{[\rho u]}\\
 \label{RHenergyconservationE}
S&=&\frac{[\rho e u +u P]}{[\rho e]}
\end{eqnarray}
\end{subequations}

The Rankine-Hugoniot jump conditions for the regularized Euler equations
\eref{NMEconservationform} are
 \begin{subequations}
\label{RHNME}
\begin{eqnarray}
\label{RHmassconservationNME}
S&=&\frac{[\rhobar \ubar -\alpha^2 (\ubar \rhobar_{xx} +\rhobar \ubar_{xx})+\alpha^2 \ubar_x \rhobar_x]}{[\rho]}\\
 \label{RHmomentumconservationNME}
S&=&\frac{[\overline{\rho u} \ubar -\alpha^2 (\ubar \overline{(\rho
u)}_{xx} +\overline{\rho u} \ubar_{xx})+\alpha^2 \ubar_x
\overline{\rho u}_x +P]}{[\rho u]}\\
 \label{RHenergyconservationNME}
S&=&\frac{[\overline{\rho e} \ubar -\alpha^2 (\ubar \overline{(\rho
e)}_{xx} +\overline{\rho e} \ubar_{xx})+\alpha^2 \ubar_x
\overline{\rho e}_x + \overline{P} \ubar -\alpha^2 (\ubar
\overline{P}_{xx} +\overline{P} \ubar_{xx})+\alpha^2 \ubar_x
\overline{P}_x]}{[\rho e]}
\end{eqnarray}
\end{subequations}

Using the exact same computations as in section
\ref{travelingwavesectionNMHE} it is a straight forward process to
show that for solutions of the form
\eref{travelingwavesolutionEuler}, the jump conditions for the Euler
equations \eref{RHfulleuler} are exactly equivalent to the jump
conditions for the regularized Euler equations \eref{RHNME}.  Thus Equation
\eref{travelingwavesolutionEuler} with values satisfying Equations
\eref{RHfulleuler} will be traveling weak solutions to the regularized Euler equations.

Additionally using the same analysis found in section
\ref{shockthicknessectionNMHE}, it can again be showing that the
thickness of the shocks for the traveling solution will decreases
linearly with $\alpha.$

\subsection{Eigenvalues}
Much like the homentropic Euler equations, the one-dimensional Euler
equations have very well defined eigenvalues $u, u \pm a$.  Again we
examine how the averaging technique has affected the eigenvalues of
the system. In this section we examine the eigenvalues of the regularized Euler equations. We begin with Equations \eref{NME} through much substitution and manipulation you can express them in primitive variable form as such:
\begin{align}
 \rho_t&+\ubar \rho_x+\rhobar u_x  =0\\
u_t&+\ubar u_x +\underbrace{\left( \frac{\overline{\rho u} -
u\rhobar}{\rho} \right)}_{\beta} u_x + P_x=0.\\
P_t&- (\gamma-1)\beta \rho u  u_x +\underbrace{\left((\gamma-1)\bar{P}+(\gamma-1)\overline{\rho e}-\frac{1}{2}(\gamma-1)\rhobar u^2\right)}_{\rho \tilde{a}^2}u_x +(\gamma \ubar -(\gamma -1)u)P_x=0.
\end{align}
The equations can then be written in vector matrix form leading to
\begin{equation} \label{matirxform}
\left[
  \begin{array}{c}
   \rho \\
     u \\
    P
  \end{array}
\right]_t +\underbrace{\left[
   \begin{array}{ccc}
     \ubar  & \rhobar & 0\\
     0 & \beta +\ubar & \frac{1}{\rho} \\
     0 & -(\gamma-1)\beta \rho u + {\rho \tilde{a}^2}  & \gamma \ubar -(\gamma -1)u
   \end{array}
 \right]}_A
 \left[
  \begin{array}{c}
    \rho \\
    u \\
    P
  \end{array}
\right]_x =0.
\end{equation}
Define the quantities $\beta$, $\kappa$, and $\tilde{a}$ as
\begin{eqnarray}
\beta&=&\frac{\overline{\rho u} - u\rhobar}{\rho} \\
\kappa&=&\ubar-u\\
\tilde{a}^2&=&\frac{\left((\gamma-1)\bar{P}+(\gamma-1)\overline{\rho e}-\frac{1}{2}(\gamma-1)\rhobar u^2\right)}{\rho}.
\end{eqnarray}
Then the eigenvalues of matrix $A$ are
\begin{eqnarray}
\lambda_0 &=&\ubar\\
\lambda_\pm&=& \frac{u+\ubar}{2} + \frac{\beta}{2} -\frac{\gamma
\kappa}{2} \pm \sqrt{\frac{\beta^2}{4}+\frac{(\gamma-1)^2\kappa^2}{4}-\frac{(\gamma-1)(\ubar+u)\beta}{2}+\tilde{a}^2}
\end{eqnarray}

The quantities $\beta$ and  $\kappa$ would appear to limit to zero
as $\alpha \to 0$, and the quantity $\tilde{a}^2$ would appear to
limit to $a^2$, the speed of sound for the Euler equations. Again the
averaging technique alters the eigenvalues, but with the original
values regained with the limit as $\alpha \to 0$.

Much like in the homentropic case the matrix $A$ can as be diagonalized and can thus be written in the form $A=Q \Lambda Q^{-1}.$ Using this we can write the regularized Euler equations in the characteristic form
\begin{equation}
\frac{\p \vb}{\p t}+ \Lambda \frac{\p \vb}{\p x}=0
\end{equation}
where
\begin{equation}
d \vb=Q^{-1} \left[\begin{array}{c}
   d \rho \\
   d u \\
   d P\\
  \end{array}\right].
\end{equation}
When diagonalizing $A$ we calculated
\begin{equation}
Q^{-1}=
\left[
  \begin{array}{ccc}
    \frac{\tilde{a}^2-(\gamma-1)\ubar \beta}{(\gamma-1)\kappa \rhobar} & 1 & \frac{-1}{\rho(\gamma-1) \kappa} \\
    0 & 1 & \frac{(\gamma-1)\kappa-\beta + \sqrt{\beta^2+(\gamma-1)^2\kappa^2-2(\gamma-1)(\ubar+u)\beta+4\tilde{a}^2}}{2\rho(\tilde{a}^2-(\gamma-1)u\beta)} \\
    0 & 1 & \frac{(\gamma-1)\kappa-\beta -\sqrt{\beta^2+(\gamma-1)^2\kappa^2-2(\gamma-1)(\ubar+u)\beta+4\tilde{a}^2}}{2\rho(\tilde{a}^2-(\gamma-1)u\beta)} \\
  \end{array}
\right].
\end{equation}
With $Q^{-1}$ containing such a complex structure we were not able to find a straightforward way of determining an analytical expression for the characteristic variables.

\subsection{Convergence to a weak solution}\label{NMEweaksolutionsection}
Section \ref{weaksolutionsection} showed that with certain
assumptions made on the solutions of the regularized Euler equations, the solutions converge to weak
solutions of the homentropic Euler equations as $\alpha \to 0$. With
similar assumptions on the regularized Euler equations, we are able to prove that
the solutions converge to weak solutions of the Euler equations.

Again the importance of showing that solutions of the regularized Euler equations
converge to weak solutions of the Euler equations is that it is
desirable that the regularized Euler equations capture the physical behavior
demonstrated in the Euler equations.  Among these is shock speed.
Convergence to a weak solution  verifies that shocks formed in the
NME equations have speeds that limit to the proper shock speeds
found in the Euler equations.

With an existence and uniqueness proof for the regularized Euler equations not yet
developed, we are forced to make several assumptions, that are
modest considering the numerical evidence presented in section
\ref{NMEnumericssection}. Assume that for every $\alpha > 0$ there
exists a solution. Beyond that assume a subsequence of those
solutions converge in $L^1_{loc}$ and the solutions are bounded
independent of $\alpha.$ The following summarizes these assumptions.
\begin{subequations}\label{NMEaasumptions}
\begin{align}
||u||_\infty < U\\
||\rho||_\infty < R\\
||e||_\infty <E\\
\lim_{\alpha \to 0}  u = \tilde{u} \mbox{ in }L^1_{loc}\\
\lim_{\alpha \to 0} \rho = \tilde{\rho} \mbox{ in }L^1_{loc}\\
\lim_{\alpha \to 0} e = \tilde{e} \mbox{ in }L^1_{loc}\\
\end{align}
\end{subequations}

With these assumptions we are able to prove that the weak solutions
to the regularized Euler equations \eref{NME} will converge to weak solutions of
the Euler equations \eref{EulerEquations} as $\alpha \to 0.$ The
examination of this claim is done with the Helmholtz filter which
has bounds already established in section \ref{weaksolutionsection}.
These bounds can be combined with Young's inequality
\cite{HunterJK:01a} and assumptions \eref{NMEaasumptions} to obtain
estimates on the filtered quantities.  The following estimates
essentially state that the filtered quantities have the same
infinity bound as the unfiltered quantities and that the first
derivatives of the filtered quantities are of order
$\frac{1}{\alpha}.$ Explicitly these estimates are
\begin{align}
||\ubar||_\infty < U\\
||\rhobar||_\infty < R\\
||\mbar||_\infty < UR\\
||\bar{e}||_\infty < E\\
||\overline{\rho e}||_\infty < ER\\
||\overline{P}||_\infty < (\gamma-1)\left(ER+\frac{1}{2}RU^2\right)\\
||\ubar_x||_\infty < \frac{1}{\alpha} U\\
||\rhobar_x||_\infty< \frac{1}{\alpha} R\\
||\mbar_x||_\infty < \frac{1}{\alpha} UR\\
||\overline{\rho e}_x||_\infty <  \frac{1}{\alpha} ER\\
||\overline{P}_x||_\infty < \frac{1}{\alpha} (\gamma-1)\left(ER+\frac{1}{2}RU^2\right)\\
\end{align}

Begin by multiplying Equations \eref{NMEconservationform} by a test
function $\phi$ and integrate over time and space.  The test
function $\phi$ has an infinite number of bounded and continuous
derivatives and in compactly supported.  After the multiplication
and integration the equations are now
\begin{subequations}
\begin{align}
\int_\mathbb{R} \int_0^T \rho_t \phi +&\left(\rhobar \ubar -\alpha^2 (\ubar \rhobar_{xx} +\rhobar \ubar_{xx})+\alpha^2 \ubar_x \rhobar_x \right)_x \phi \, dt \, dx=0\\
\int_\mathbb{R} \int_0^T (\rho u)_t \phi +&\left(\mbar \ubar -\alpha^2 (\ubar \overline{(\rho u)}_{xx} +\mbar \ubar_{xx})+\alpha^2 \ubar_x \mbar_x \right) +P)_x \phi \, dt \, dx=0\\
\int_\mathbb{R} \int_0^T  (\rho e)_t  \phi+&\left[\overline{\rho e} \ubar -\alpha^2 (\ubar \overline{(\rho e)}_{xx} +\overline{\rho e} \ubar_{xx})+\alpha^2 \ubar_x \overline{\rho e}_x + \right. \nonumber \\
                  +&\left.\overline{P} \ubar -\alpha^2 (\ubar \overline{P}_{xx} +\overline{P} \ubar_{xx})+\alpha^2 \ubar_x \overline{P}_x \right]_x \phi \, dt \,
                  dx=0.
\end{align}
\end{subequations}
Integrate by parts to obtain
\begin{subequations}
\begin{align}
\int_\mathbb{R} \int_0^T \rho \phi_t +\left(\rhobar \ubar \right) \phi_x \, dt \, dx=&\int_\mathbb{R} \int_0^T \left(  \alpha^2(\ubar \rhobar_{xx} +\rhobar \ubar_{xx})- \alpha^2\ubar_x \rhobar_x \right) \phi_x \, dt \, dx\\
\int_\mathbb{R} \int_0^T (\rho u) \phi_t  +\left(\mbar \ubar +P \right) \phi_x \, dt \, dx =& \int_\mathbb{R} \int_0^T   \left( \alpha^2(\ubar \overline{(\rho u)}_{xx} +\mbar \ubar_{xx})-  \alpha^2 \ubar_x \mbar_x \right) ) \phi_x \, dt \, dx\\
\int_\mathbb{R} \int_0^T  (\rho e) \phi_t +\left(\overline{\rho e} \ubar +\overline{P} \ubar \right) \phi_x \, dt \, dx=&\int_\mathbb{R} \int_0^T   \left( \alpha^2(\ubar \overline{(\rho e)}_{xx} +\overline{\rho e} \ubar_{xx})-  \alpha^2 \ubar_x \overline{\rho e}_x \right) ) \phi_x \, dt \, dx \nonumber \\
+&\int_\mathbb{R} \int_0^T   \left( \alpha^2(\ubar \overline{P}_{xx}
+\overline{P} \ubar_{xx})-  \alpha^2 \ubar_x \overline{P}_x \right)
) \phi_x \, dt \, dx.
\end{align}
\end{subequations}
Clearly if the right hand side of the above equations limits to $0$
as $\alpha \to 0$ then convergence to a weak solution is proven.
Since the calculations are all but identical, the reader is referred
to section \ref{weaksolutionsection} to prove that the right hand
side of the equations do limit to $0$. With the right hand side
limiting to $0$ we find that $\rho$, $u$, and $e$ are, in fact,
limiting to a weak solution of the Euler equations.

\subsection{Numerics}\label{NMEnumericssection}
The numerical simulations for the regularized Euler equations are the same as
those discussed in section \ref{numericssection} with the addition
of the energy equation. The equations simulated here are
\begin{subequations}
\label{NMEsmoothonlyform}
\begin{align}
\rhobar_t +  \left(\rhobar \ubar               \right)_x      =& -3\alpha^2 \overline{(\ubar_x \rhobar_x)_x}\\
\mbar_t   +  \left(\mbar   \ubar     +\bar{P}   \right)_x      =& -3\alpha^2 \overline{(\ubar_x \mbar_x)_x}\\
\overline{\rho e}_t   +  \left(\overline{\rho e}   \ubar     +\ubar
\bar{P}   \right)_x      =& -3\alpha^2 \overline{\left(\ubar_x
\left( \overline{\rho e}_x+ \overline{P}_x \right)\right)_x}.
\end{align}
\end{subequations}

The same pseudo-spectral method is used to solve Equations
\eref{NMEsmoothonlyform}.  Again an adaptive Runge-Kutta is used to
advance in time with all spatial derivatives and the inversion of
the Helmholtz operator conducted in the Fourier domain.  The
simulations were conducted with $2^{14}=65536$ grid points on the
domain $[0, 2\pi]$ for approximately 10,000 time steps.  Numerical runs were conducted for values
$\alpha=0.10, 0.09, ..., 0.02, 0.01.$  The same long term
instability was found in these numerical simulations and were again
controlled by setting wave modes higher than $\frac{N}{3}$ to zero
every 200 time steps.

\subsection{Numerical results}\label{NMEnumericalresultssection}
Again the preferred example problem is the shock tube or Riemann
problem.   This is chosen for its demonstration of expansion waves,
shocks, and contact surfaces. The example problem is one of the
classic Sod test problems \cite{SodGA:78a} where the initial
conditions are
\begin{subequations}\label{NMEexample}
\begin{align}
u_0(x)&=&0\\
\rho_0(x)&=&\left\{\begin{array}{ll}
1 & 0<x\leq\pi \\
0.125  &  \pi <x\leq 2\pi
\end{array}\right.\\
P_0(x)&=&\left\{\begin{array}{ll}
10 & 0<x\leq\pi \\
1  &  \pi <x\leq 2\pi
\end{array}\right.
\end{align}
\end{subequations}
These initial conditions were chosen as they produce an expansion
wave, a contact surface, and a shock, the three classical behaviors
of the Riemann problem in the Euler equations. As with the
homentropic Euler equations the initial conditions are twice
filtered.

 The numerical method that we chose forces periodic boundary conditions.
Thus for the shock tube problem we chose, errors begin to propagate
from the edges as time progresses.  While the simulations were
conducted on $[0, 2 \pi]$ we only consider the results valid for
$[\frac{\pi}{2},\frac{3 \pi}{2}]$ up to time $t=0.25$.

\begin{figure}[!ht]
\begin{center}
\begin{minipage}{0.29\linewidth} \begin{center}
  \includegraphics[width=\linewidth]{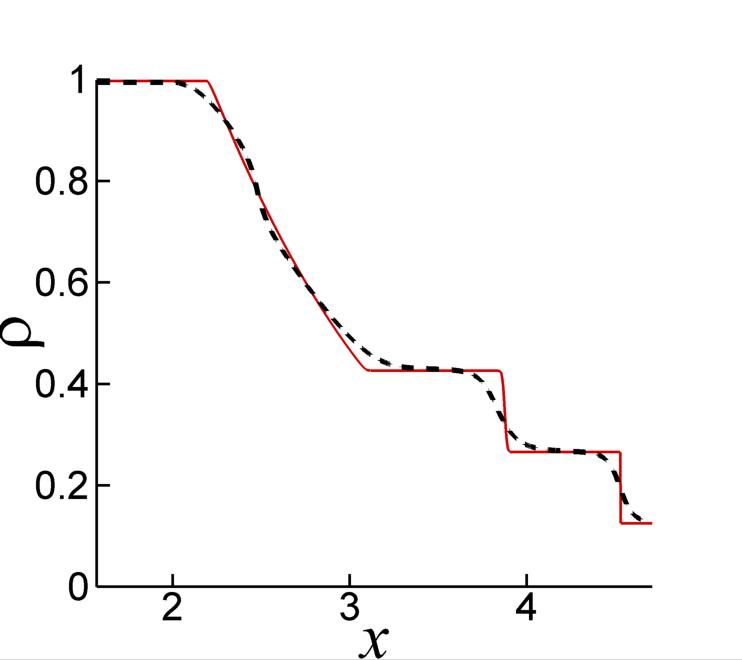}
\end{center} \end{minipage}
\begin{minipage}{0.29\linewidth} \begin{center}
  \includegraphics[width=\linewidth]{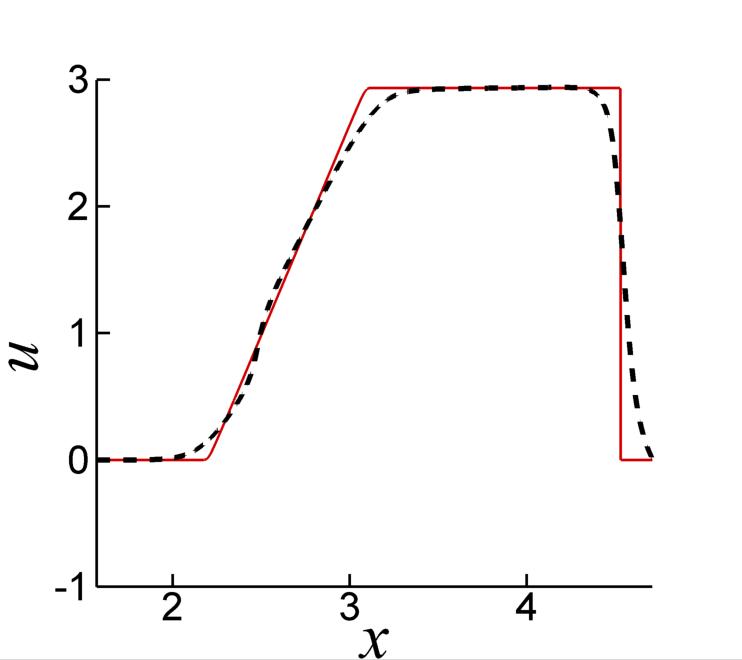}
\end{center} \end{minipage}
\begin{minipage}{0.29\linewidth} \begin{center}
  \includegraphics[width=\linewidth]{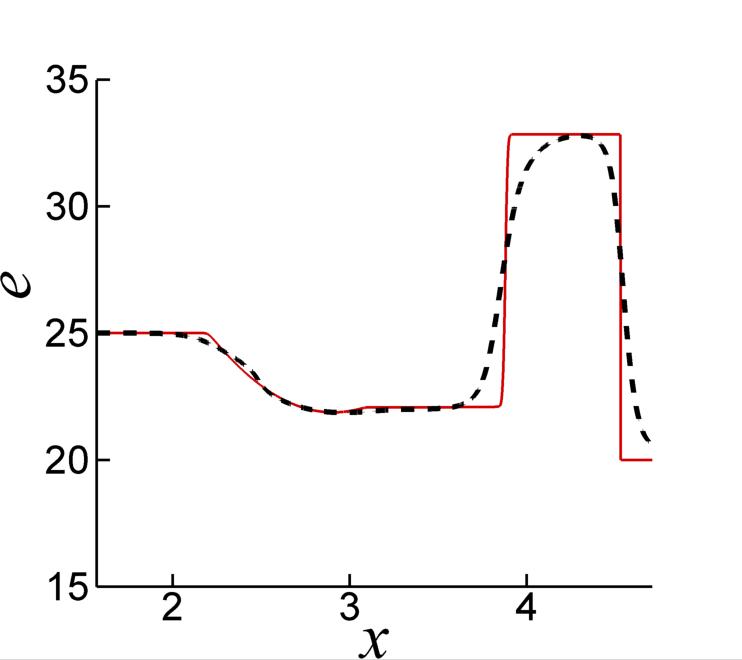}
\end{center} \end{minipage}\\
\begin{minipage}{0.29\linewidth}\begin{center} (a) \end{center}\end{minipage}
\begin{minipage}{0.29\linewidth}\begin{center} (b) \end{center}\end{minipage}
\begin{minipage}{0.29\linewidth}\begin{center} (c) \end{center}\end{minipage}\\
\caption{This figure shows a numerical simulation of the regularized Euler equations (dashed line) plotted against the solution to the  Euler
equations (solid line). Here the value of $\alpha=0.05$ at time
$t=0.25$. In the figures, it is clear that the regularized Euler equations are
capturing both the expansion wave, contact surface, and shock
behavior.  (a) The density. (b) The velocity. (c) The energy.}
\label{NMEalpha5}
\end{center}
\end{figure}

\begin{figure}[!ht]
\begin{center}
\begin{minipage}{0.29\linewidth} \begin{center}
  \includegraphics[width=\linewidth]{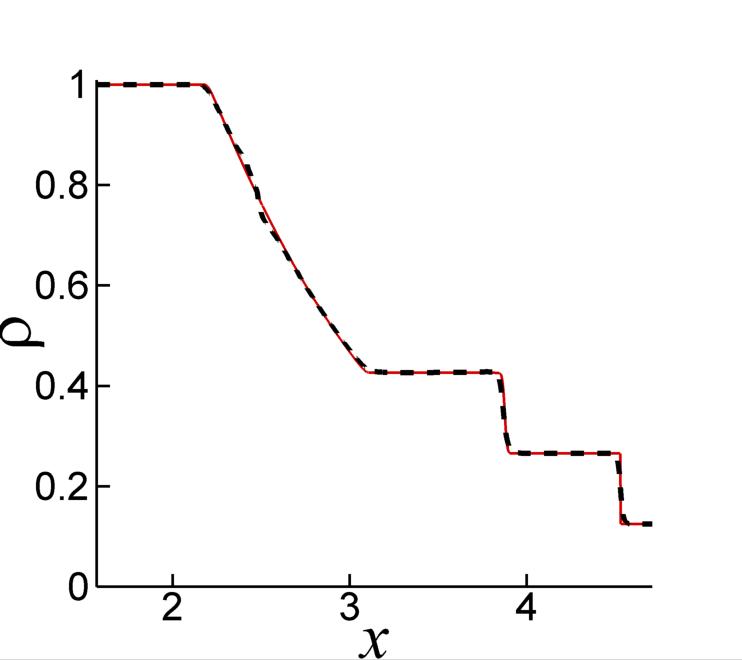}
\end{center} \end{minipage}
\begin{minipage}{0.29\linewidth} \begin{center}
  \includegraphics[width=\linewidth]{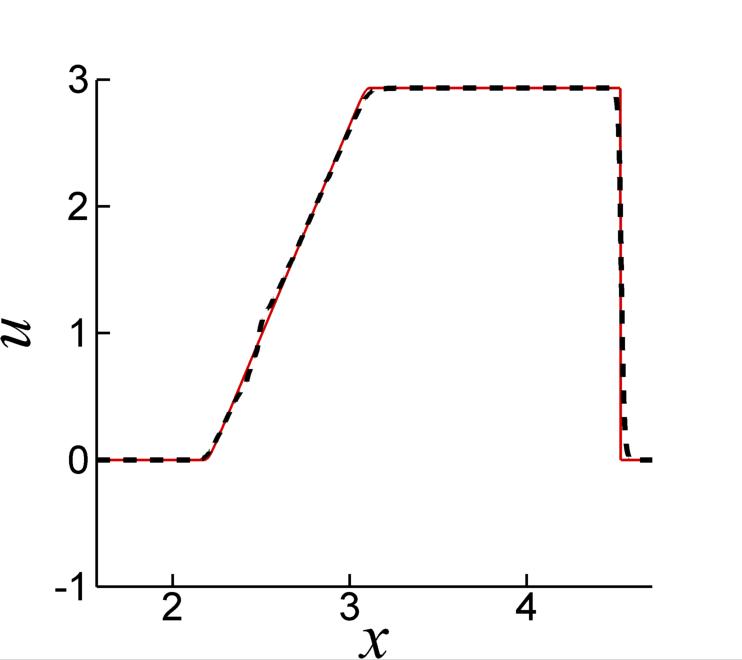}
\end{center} \end{minipage}
\begin{minipage}{0.29\linewidth} \begin{center}
  \includegraphics[width=\linewidth]{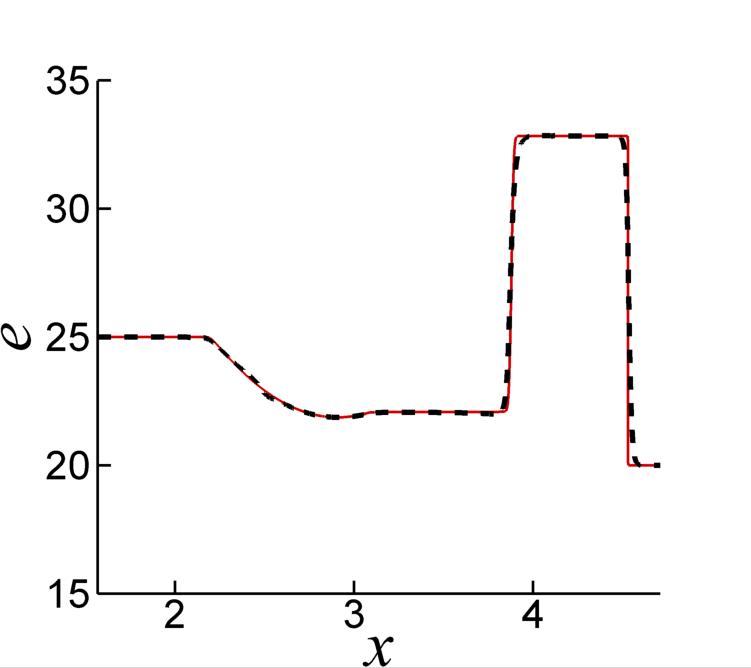}
\end{center} \end{minipage}\\
\begin{minipage}{0.29\linewidth}\begin{center} (a) \end{center}\end{minipage}
\begin{minipage}{0.29\linewidth}\begin{center} (b) \end{center}\end{minipage}
\begin{minipage}{0.29\linewidth}\begin{center} (c) \end{center}\end{minipage}\\
\caption{This figure shows a numerical simulation of the regularized Euler equations (dashed line) plotted against the solution to the  Euler
equations (solid line). Here the value of $\alpha=0.01$ at time
$t=0.25$.  In the figures, it is clear that the regularized Euler equations are
capturing both the expansion wave, contact surface, and shock
behavior.  With the lower value of $\alpha$ the fit is much closer.
(a) The density. (b) The velocity. (c) The energy.}
\label{NMEalpha1}
\end{center}
\end{figure}

Figures \ref{NMEalpha5} and  \ref{NMEalpha1} show simulations of the
regularized Euler equations for two different values of $\alpha$ plotted against
the solution to the Euler equations.  The expansion wave, contact
surface, and the shock are all captured in the behavior of the regularized Euler equations.  For the smaller value of $\alpha$, the behavior of the
regularized Euler equations matches the Euler equations more closely.

\begin{figure}[!ht]
\begin{center}
  \includegraphics[width=.5\linewidth]{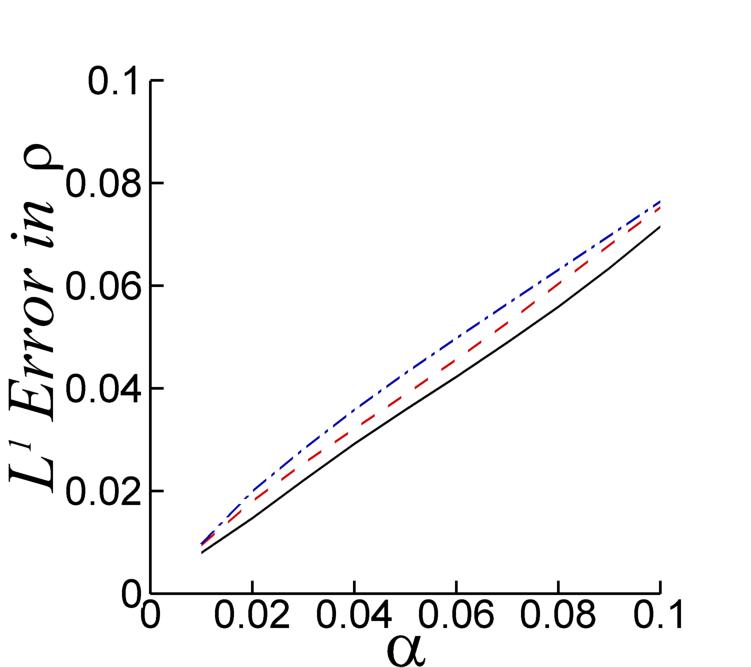}
\caption{This figure displays the difference between the density in
solutions of the regularized Euler equations and the solution of the Euler
equations in the $L^1$ norm as $\alpha \to 0$.  As $\alpha \to 0$
the difference in the solutions also approaches zero.  The
measurements were taken for $\alpha=0.01, 0.02, ..., 0.1$ at times
$t=0.05$, \solid, $t=0.15$ \dashed, and $t=0.25$ \dashdot.}
\label{NMEl1rhoerror}
\end{center}
\end{figure}

\begin{figure}[!ht]
\begin{center}
  \includegraphics[width=.5\linewidth]{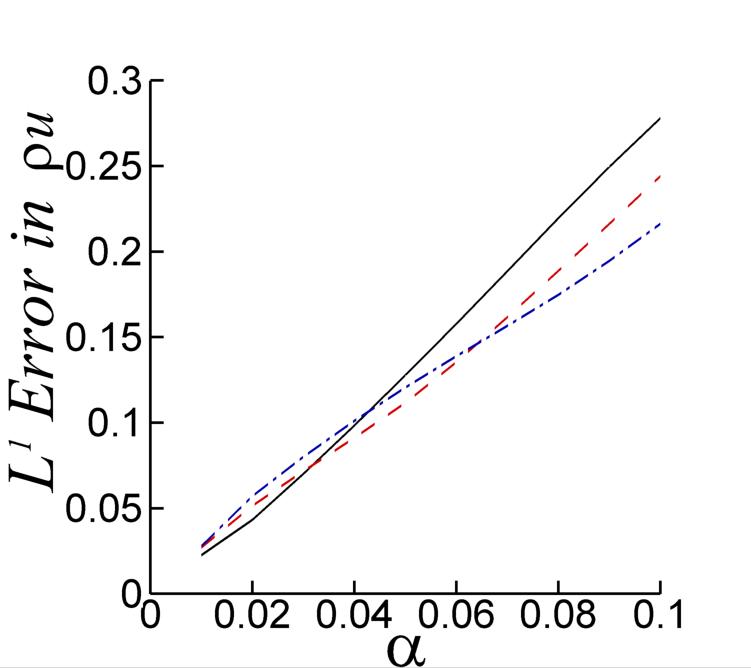}
\caption{This figure displays the difference between the momentum in
solutions of the regularized Euler equations and the solution of the Euler
equations in the $L^1$ norm as $\alpha \to 0$.  As $\alpha \to 0$
the difference in the solutions also approaches zero.  The
measurements were taken for $\alpha=0.01, 0.02, ..., 0.1$ at times
$t=0.05$, \solid, $t=0.15$ \dashed, and $t=0.25$ \dashdot.}
\label{NMEl1merror}
\end{center}
\end{figure}

\begin{figure}[!ht]
\begin{center}
  \includegraphics[width=.5\linewidth]{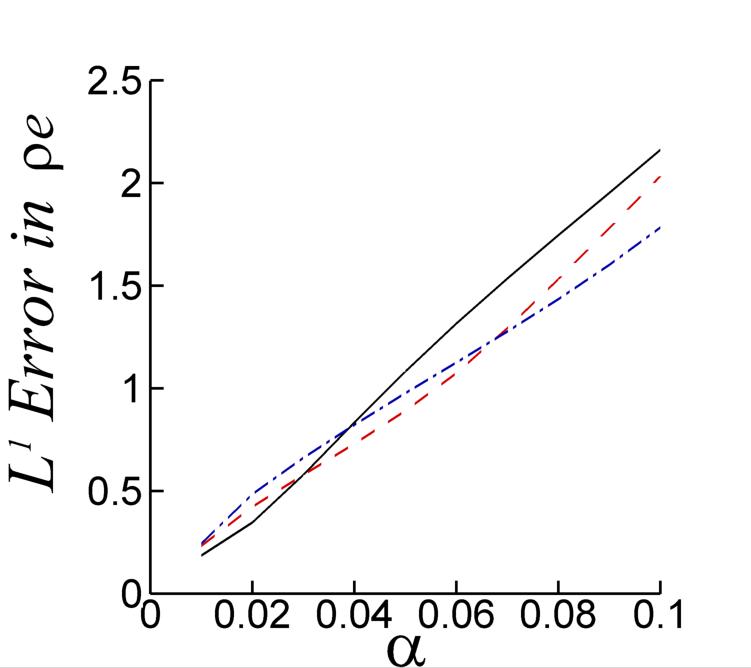}
\caption{This figure displays the difference between the energy in
solutions of the regularized Euler equations and the solution of the Euler
equations in the $L^1$ norm as $\alpha \to 0$.  As $\alpha \to 0$
the difference in the solutions also approaches zero.  The
measurements were taken for $\alpha=0.01, 0.02, ..., 0.1$ at times
$t=0.05$, \solid, $t=0.15$ \dashed, and $t=0.25$ \dashdot.}
\label{NMEl1peerror}
\end{center}
\end{figure}

As before we check the convergence of the solutions of the regularized Euler equations to the solution of the Euler equations as $\alpha \to 0$.
Figures \ref{NMEl1rhoerror}, \ref{NMEl1merror}, and
\ref{NMEl1peerror} show that as $\alpha \to 0$ the error in the
$L^1$ norm appears to be approaching zero for the example problem.
This suggests that the solutions of the regularized Euler equations converge to
the solutions of the Euler equations.

\subsection{Kinetic energy rates}
As in section \ref{kineticenergysectionNMHE}, we examine the kinetic energy for the shock tube problem.  Again for the shock tube problem, the solution to the Euler equations are self similar, depending only on the variable $\frac{x}{t}$ and thus the kinetic energy changes linearly in time. For the Euler equations, we examine the kinetic energy $\frac{1}{2} \rho u^2$ and for the regularized Euler equations we examine an unfiltered kinetic energy $\frac{1}{2} \rho u^2$ and a filtered kinetic energy $\frac{1}{2} \rhobar \ubar^2.$  Other filtered kinetic energies were also considered, but as in the homentropic case, for this example problem, the differences between them and $\frac{1}{2} \rhobar \ubar^2$ were negligible.

\begin{figure}[!ht]
\begin{center}
\begin{minipage}{0.48\linewidth} \begin{center}
  \includegraphics[width=.9\linewidth]{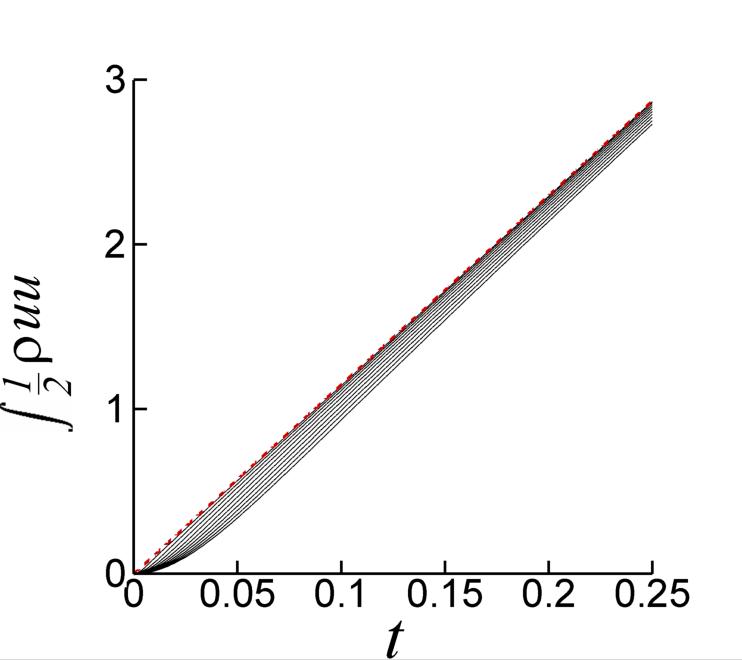}
\end{center} \end{minipage}
\begin{minipage}{0.48\linewidth} \begin{center}
  \includegraphics[width=.9\linewidth]{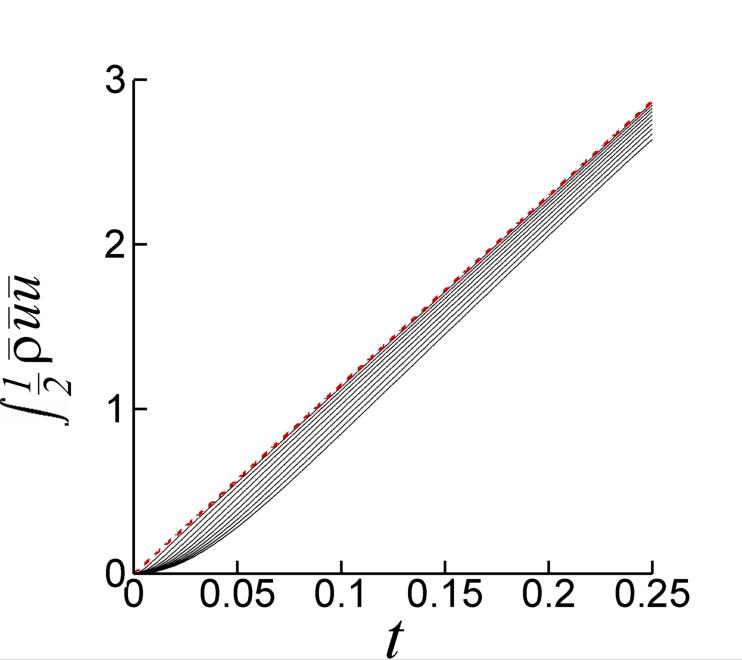}
\end{center} \end{minipage}\\
\begin{minipage}{0.48\linewidth}\begin{center} (a) \end{center} \end{minipage}
\begin{minipage}{0.48\linewidth}\begin{center} (b) \end{center}
\end{minipage}
\caption{The kinetic energy of the Euler equations and the regularized Euler equations. The energy for the true solution of the Euler equations is shown as a dashed red line.  The simulation of the regularized Euler equations for $\alpha=0.01, 0.02, ..., 0.1$ are shown as solid black lines.  The bottommost line represents $\alpha=0.1$.  As $\alpha$ decreases the energy approaches the energy of the Euler equations.  (a) These are plots of the unfiltered kinetic energies, $\frac{1}{2} \rho u^2$.  (b)  These are plots of the filtered kinetic energies, $\frac{1}{2} \rhobar \ubar^2$.  When examined the plots of $\frac{1}{2} \mbar \ubar$ and $\frac{\mbar^2}{2 \rhobar}$ were identical to this one. } \label{kinecticenergyfigureNME}
\end{center}
\end{figure}

Figure \ref{kinecticenergyfigureNME} shows how the kinetic energies for the regularized Euler equations behave for various values of $\alpha.$  Again after a brief period, the energies seem to vary linearly with time.  We attribute this brief period to the averaging of the initial conditions.  As $\alpha$ decreases the energies for the regularized Euler equations approach the energy of the Euler equations, as we would expect if the solutions are converging as $\alpha \to 0$.

\section{Conclusion}
Using the convectively filtered Burgers (CFB) equations as
inspiration we have developed a new averaging technique with the
intent of regularizing both shocks and turbulence simultaneously.
This paper examines primarily the shocks regularization aspect of
the technique.  We discussed the physical motivation for the method and then derived
a general technique to be used on conservation laws. It was then
established that this technique, when applied to conservation laws,
would preserve the original conservative properties.

The remainder of the paper then examined the effects when this
method was applied to the homentropic Euler and Euler equations. The
results show much promise.  It was found that with the Helmholtz
filter that both the regularized homentropic Euler and regularized Euler equations can be rewritten in conservation form, reiterating that the original conservative
properties are preserved.  For both sets of equations we were able
to find traveling shock solutions, where the Rankine-Hugoniot jump
conditions for the modified equations reduced to the same jump
conditions of the original equations.  For both sets of equations it
was proven that as the filtering is decreased, $\alpha \to 0$, the
solutions will converge to weak solutions of the original equations.
Both these results show that the proper shock speeds of the original
equations will be preserved with the averaging techniques.

Numerical simulations were run on both sets of equations for the
shock tube problem.  These simulations demonstrated that the
modified equations mimic the behavior of the original equations. The regularized homentropic Euler equations captured both the expansion wave and the shock front
behavior, while the regularized Euler equations captured the expansion wave,
contact surface,  and the shock front.  The solutions appeared
regularized meaning that they were continuous and smooth.
Furthermore, as $\alpha \to 0$ these solutions were seen to be
converging to the solutions of the original equations.

There is still more work to be done regarding these equations.  It
would be beneficial to establish more theory regarding both sets of
equations. Specifically existence proofs would be beneficial. It
would be interesting to see if the either sets of equations possess
a Hamiltonian structure.

The regularized homentropic Euler and regularized Euler equations are showing promise as a new
regularization method based on the preliminary examination
considered here.  From these results we believe that this averaging
technique leads to a regularization of the homentropic Euler and
Euler equations that is capable of capturing the relevant behavior
of the equations, at least for short time simulations.  With future
work we hope to establish this more thoroughly and extend this
technique into higher dimensions.

\section{Acknowledgments}
The research in this paper was partially supported by the AFOSR
contract FA9550-05-1-0334.  We would also like to thanks Dr. Keith Julien for his advice in the numerical simulations.

\bibliography{../RefA1}
\bibliographystyle{unsrt}

\end{document}